\documentclass[twocolumn]{aastex631}

\usepackage{amsmath}
\usepackage{threeparttable}

\newcommand{\toacode}{\textsc{toa\_sp}}

\shorttitle{Direct Single-Pulse Timing}
\shortauthors{Zhang et al.}

\begin{document}

\title{TOA\_SP: A Multi-Strategy Framework for Single-Pulse Timing}

\author{Songbo Zhang}
\affiliation{Purple Mountain Observatory, Chinese Academy of Sciences, Nanjing 210023, China}

\author{Xuan Yang}
\affiliation{Purple Mountain Observatory, Chinese Academy of Sciences, Nanjing 210023, China}

\correspondingauthor{Songbo Zhang, Xuan Yang}
\email{sbzhang@pmo.ac.cn, yangxuan@pmo.ac.cn}

\begin{abstract}
Precision pulsar timing typically relies on the stability of average pulse profiles, enabling time-of-arrival (TOA) estimation through template cross-correlation. This assumption breaks down for highly variable radio sources such as Rotating Radio Transients (RRATs) and fast radio bursts (FRBs), where individual pulses could exhibit strong variability in morphology and amplitude, and no single averaged profile may represent the underlying emission process.
We present \textsc{toa\_sp}, an open-source Python package for extracting TOAs directly from \textsc{PSRFITS} search-mode data without requiring profile folding into a stable template. The framework implements a suite of complementary single-pulse timing strategies, including parametric profile fitting, non-parametric estimators, and adaptive sub-band and time-resolution optimisation, together with empirical diagnostics for assessing model consistency.
We apply \textsc{toa\_sp} to 688 single pulses from a 3-hour FAST observation of RRAT~J1913+1330. The resulting TOAs residual achieve a weighted RMS residual of 1.33\,ms, a 24\% improvement over a standard template-based \textsc{PSRCHIVE} pipeline, while retaining all pulses without statistical outlier rejection. A set of bright FRB 20220529 bursts provides a controlled test of the framework across regimes of increasing pulse complexity, revealing frequency-dependent substructure not captured by band-integrated profiles.
We introduce an empirical convergence diagnostic that identifies well-constrained pulses and guides the transition between parametric and non-parametric regimes. Full multi-strategy processing of 688 pulses requires approximately 7.6\,s per pulse on a 10-thread CPU. The package is publicly available via \texttt{pip install toa\_sp}.
\end{abstract}


\section{Introduction} \label{sec:intro}

Pulsar timing measures pulse arrival times to extraordinary precision. For the most stable millisecond pulsars (MSPs), timing residuals reach $\sim$100\,ns \citep{Verbiest16,Antoniadis22}, enabling pulsar timing arrays (PTAs) to search for the nanohertz gravitational-wave background \citep{Agazie23,Antoniadis23,Reardon23,Xu23}. The standard pipeline folds raw search-mode data modulo the pulsar period to form an integrated profile, then cross-correlates with a high-S/N template to extract the TOA \citep{Taylor89}. Packages such as \textsc{dspsr} \citep{vanStraten11} and \textsc{psrchive} \citep{Hotan04,vanStraten12} have optimized this workflow over decades. In practice, standard timing workflows using these pipelines are optimized for \emph{fold-mode} data, where the raw voltages have already been folded at the observed pulse period. FAST, in contrast, predominantly records \emph{search-mode} PSRFITS data, preserving full time--frequency resolution \citep{Jiang20}. This format is more flexible for single-pulse studies but requires tools that operate directly on un-folded data.

The standard pulsar timing pipeline assumes profile stability, whereby averaging over many rotations produces a stable integrated pulse profile \citep{Lorimer04}. While this assumption generally holds for most pulsars, mode-changing pulsars \citep{Wang07} alternate between two or more quasi-stable profile states, requiring separate timing templates. At the extreme, Rotating Radio Transients (RRATs) \citep{McLaughlin06} might emit detectable pulses during fewer than 10\% of rotations, and their active-phase pulses can vary dramatically in width, peak flux density, and morphology from one rotation to the next \citep{Keane08,McLaughlin09,Bhattacharyya18,Zhang24}. For highly variable RRATs \citep{Zhang24}, the integrated profile averages unrelated emission components from different active phases and therefore no longer represents a physically meaningful timing reference.

The conventional folding-based timing pipeline has several important limitations when applied to highly variable sources. First, variable pulse shapes and phases prevent the integrated profile from representing any physical emission component; template-matched TOAs become weighted averages of unrelated features. Second, dense sub-structure in bright pulses is smeared into an unresolved superposition, causing the measured TOA to depend on the component that dominates the template. Third, folding multi-hour FAST observations through \textsc{dspsr} can take days \citep{Zhang24_highB}, creating a computational bottleneck for large surveys targeting sources with low event rates. Furthermore, FRBs present a more fundamental limitation: without a known rotation period, folding is impossible, and their single pulses must be timed directly from search-mode data.

Our alternative approach extracts TOAs directly from individual single pulses, bypassing folding entirely. Standard search methods, such as \texttt{heimdall}\footnote{\url{https://sourceforge.net/projects/heimdall-astro/}} and \texttt{presto}\footnote{\url{http://www.cv.nrao.edu/~sransom/presto/}} \citep{Ransom01}, first detect pulses and determine their DMs and approximate arrival times \citep{Zhang24}. Our framework then performs high-precision TOA extraction. This approach has been used for individual FRB bursts \citep{Nimmo22} and for single-pulse RRAT timing \citep{Zhang24}, but no systematic study has examined which fitting strategies yield reliable TOAs across the full parameter space of S/N and profile complexity. No unified software framework yet exists.

We present \toacode{}, a Python package that fills this gap. It implements nine complementary TOA estimation strategies, including parametric Gaussian-, EMG-, and Voigt-based fitting, non-parametric leading-edge, center-of-mass, peak, and shapelet methods, together with MCMC uncertainty estimation, automatic time-resolution optimization, and sub-band cross-validation. An empirical convergence diagnostic identifies pulses that require more sophisticated modeling.

Section~\ref{sec:methods} describes the methodology, including the TOA strategies, sub-band cross-validation, and the AICc-based model selection. Section~\ref{sec:sample} presents the source sample. Section~\ref{sec:results} demonstrates the framework on 688 single pulses from RRAT~J1913$+$1330 \citep{Zhang24} to show that direct timing outperforms conventional templates for highly variable sources, and then validates the multi-strategy diagnostics using three bright bursts from FRB~20220529 \citep{Li26}. Section~\ref{sec:discussion} discusses the implications, including the convergence diagnostic, outlier rejection philosophy, and limitations. Section~\ref{sec:conclusion} summarizes the main results and outlines recommended usage of the \toacode{} framework.

\section{Methodology} \label{sec:methods}

\subsection{Pipeline Overview}

The \toacode{} pipeline operates directly on PSRFITS search-mode sub-integration data without folding. Unlike traditional pipelines that require all sub-integrations from an observation to be summed prior to processing, \toacode{} accepts individual sub-integration files for per-file analysis. The processing consists of five steps:

\begin{enumerate}
\item {\bf Read and unpack.} Sub-integration data arrays are extracted using \textsc{astropy} \citep{astropy22}, and 8-bit packed samples are unpacked into channelized time series using vectorized \textsc{numpy} operations \citep{numpy20}.
\item {\bf Dedispersion and binning.} Incoherent dedispersion at the nominal DM aligns all frequency channels to the top of the band. The data are optionally binned in frequency (factor \texttt{bf}) and time (factor \texttt{bs}) to achieve the desired resolution.
\item {\bf RFI excision.} Broadband RFI is mitigated by subtracting a zero-DM time series that captures impulsive broadband interference commonly observed at FAST. Narrow-band RFI is then identified and flagged using iterative 3-$\sigma$ thresholds applied per channel; flagged channels are replaced by their in-band mean.
\item {\bf Baseline subtraction.} A running median filter (window length $\sim$50 times the expected pulse width) removes slowly varying baseline structure.
\item {\bf Pulse fitting and TOA extraction.} The frequency-averaged profile (optionally restricted to a user-defined band; Section~\ref{sec:freq_range}) is optionally rebinned to an optimal time resolution by maximizing the peak signal-to-noise ratio (S/N, Section~\ref{sec:auto_res}). The resulting profile is then fit using parametric models (Gaussian, EMG, Voigt, or shapelet; Section~\ref{sec:fitting}) plus a constant offset. TOAs are extracted using one of nine strategies (Section~\ref{sec:strategies}).
\end{enumerate}

\subsection{Profile Fitting Framework} \label{sec:fitting}

The default model is an $N$-component Gaussian profile with a constant baseline,
\begin{equation}
S(t) = \sum_{i=1}^{N} A_i \exp\left[-\frac{(t-\mu_i)^2}{2\sigma_i^2}\right] + C,
\label{eq:ngauss}
\end{equation}
where $A_i$, $\mu_i$, and $\sigma_i$ denote the amplitude, centroid, and width of each component, and $C$ is a constant offset. Negative amplitudes are allowed to accommodate absorption-like features, although emission components are expected to satisfy $A_i > 0$.

For asymmetric or scattered pulse profiles, two additional models are supported (Section~\ref{sec:emg_voigt}): the exponentially modified Gaussian (EMG), obtained via convolution $G(t;\mu,\sigma) \otimes \exp(-t/\tau)\Theta(t)$, and the Voigt profile, $G(t;\mu,\sigma) \otimes L(t;\gamma)$. These models introduce additional parameters $\tau_i$ and $\gamma_i$, respectively. Model selection is controlled via \texttt{--profile\_model}.

Fitting is performed using \textsc{scipy.optimize.curve\_fit} \citep{scipy20} with Trust Region Reflective bounds \citep{Branch99}. The width parameter is constrained to $\sigma_i \in [10^{-6}, 0.5\Delta T]$, where $\Delta T$ is the fitting window size, and component centroids are restricted within the window. For single-component fits ($N=1$), a two-stage procedure is used: an initial wide-window fit estimates the pulse width, followed by a refined fit within a $\pm 3\,\mathrm{FWHM}$ window.

For multi-component models ($N \geq 2$), initial guesses are obtained using \textsc{scipy.signal.find\_peaks}, with adaptive thresholds defined relative to the off-pulse RMS ($\sigma_{\rm off}$). Peak positions determine initial centroids, amplitudes are initialized from local maxima above the median baseline, and initial widths are set to $w_0/6$.

\subsection{TOA Extraction Strategies} \label{sec:strategies}

We implement nine strategies for extracting a single TOA from the fitted model or the profile directly. Table~\ref{tab:strategies} summarizes them.

\begin{table*}
\caption{TOA extraction strategies implemented in \toacode{}. All parametric strategies support Gaussian, EMG, and Voigt profile models (selectable via \texttt{--profile\_model}).}
\label{tab:strategies}
\centering
\begin{tabular}{lll}
\hline
Strategy & Type & Description \\
\hline
\texttt{single}       & Parametric & Single Gaussian, two-pass fit; TOA = $\mu_1$ \\
\texttt{highest}      & Parametric & Brightest component: $\arg\max_i |A_i|$ \\
\texttt{error\_weighted} & Parametric & Inverse-variance weighted: $\sum(\mu_i/\sigma_i^2)/\sum(1/\sigma_i^2)$ \\
\texttt{first\_peak}  & Parametric & Earliest component with S/N ${}>5$; skipped if none \\
\texttt{weighted}     & Parametric & Amplitude-weighted: $\sum|A_i|\mu_i/\sum|A_i|$ (deprecated) \\
\texttt{leading\_edge}& Non-parametric & Half-max crossing + rise-time correction to peak level \\
\texttt{center\_of\_mass} & Non-parametric & First moment above $3\sigma_{\rm off}$ \\
\texttt{peak}           & Non-parametric & Profile maximum with quadratic sub-bin interpolation \\
\texttt{shapelet}       & Non-parametric & Shapelet decomposition; TOA $=$ peak of reconstructed model \\
\hline
\end{tabular}
\end{table*}

\subsubsection{Parametric vs.\ Non-parametric Strategies}

The nine TOA strategies fall into two categories, distinguished by whether the pulse profile is assumed to follow a prescribed functional form. Parametric strategies model the profile as a sum of $N$ Gaussian (or EMG/Voigt) components, each parameterized by $\{A_i, \mu_i, \sigma_i\}$. In this case, the functional form is fixed and the fit determines only the parameter values. This approach yields high statistical precision when the true pulse shape is well described by the model, with TOA uncertainties typically scaling as $\sigma/({\rm S/N}\sqrt{N_{\rm eff}})$. However, the result is only as reliable as the underlying model assumptions. When the pulse contains overlapping sub-components that are not well described by a small number of Gaussians, the fit becomes degenerate, and the inferred TOA may change discontinuously depending on the chosen decomposition (Section~\ref{sec:results}). 

Non-parametric strategies make no explicit assumptions about the pulse functional form. Instead, they estimate the TOA directly from geometric properties of the observed profile (e.g., the leading edge or flux-weighted centroid), or from flexible basis representations whose complexity adapts to the data. These methods are less sensitive to model misspecification but generally yield larger uncertainties, particularly in low S/N or highly structured pulses. They can instead be viewed as addressing a different question: \emph{given the observed profile, what is the most robust estimate of the arrival time?}

\subsubsection{Parametric Strategies} \label{sec:param_strategies}

All parametric strategies operate on the output of an $N$-component Gaussian fit (Eq.~\ref{eq:ngauss}; EMG or Voigt models are selected via \texttt{--profile\_model}). They differ only in how a single TOA is extracted from the fitted components.

\texttt{single} ($N=1$) is the simplest parametric strategy: a single Gaussian is fitted using the two-pass procedure described in Section~\ref{sec:fitting}, and the centroid $\mu_1$ is taken as the TOA.

\texttt{highest} selects the component with the largest absolute amplitude, corresponding to the dominant emission feature in the fitted decomposition.

\texttt{error\_weighted} computes a weighted average of component centroids using the inverse variance of their formal uncertainties ($1/\sigma_{\mu,i}^2$), thereby down-weighting poorly constrained components.

\texttt{first\_peak} selects the earliest statistically significant component based on signal-to-noise and width constraints, thereby reducing sensitivity to spurious components arising from noise or fitting degeneracies.

\texttt{weighted} uses amplitude-weighted averaging of component centroids. This strategy is included for completeness but is sensitive to low-S/N or poorly constrained components, and may produce biased TOAs in such cases (Section~\ref{sec:results}).

\subsubsection{Non-parametric Strategies} \label{sec:nonpar_strategies}

The four non-parametric strategies operate directly on the 1D pulse profile without reference to a prescribed component model.

{\bf Leading edge (\texttt{leading\_edge}).}
The TOA is defined from the rising edge of the pulse. The algorithm locates the half-maximum crossing on the leading flank (50\% of $S_{\rm peak}-S_{\rm baseline}$, with linear interpolation for sub-bin precision), and maps this point to the pulse peak using the measured rise time. This construction reduces sensitivity to pulse width variations while retaining robustness to scattering-induced trailing structure. For symmetric profiles, the estimator reduces to the peak position. The TOA uncertainty is estimated approximately as $t_{\rm rise}/{\rm S/N}$.

{\bf Peak (\texttt{peak}).}
The simplest estimator: the TOA is taken as the location of the maximum of the profile, refined using a quadratic interpolation over three adjacent bins. This method is fully model-independent and computationally inexpensive, but is sensitive to noise fluctuations and can be biased in the presence of asymmetric scattering tails. It serves as a baseline reference estimator.

{\bf Center of mass (\texttt{center\_of\_mass}).}
The TOA is defined as the first moment of the profile above a baseline threshold:
\[
\mu = \frac{\int t\,[S(t)-S_{\rm th}]\,dt}{\int [S(t)-S_{\rm th}]\,dt}, \quad S_{\rm th}=3\sigma_{\rm off}.
\]
This thresholding reduces the influence of off-pulse noise. The method is simple and parameter-free but can be biased toward later arrival times in the presence of extended scattering tails.

{\bf Shapelet decomposition (\texttt{shapelet}).}
We also implement a shapelet-based representation following \citet{Refregier03,Ashton20}, in which the pulse profile is expanded in a set of Hermite functions modulated by a Gaussian envelope, with an additional low-order polynomial baseline:
\begin{equation}
S(t) = \sum_{i=0}^{n_s} C_i\,H_i\!\left(\frac{t-\tau}{\beta}\right)
       \exp\!\left[-\left(\frac{t-\tau}{\beta}\right)^2\right]
     + \sum_{j=0}^{n_p} B_j (t - t_{\rm mid})^j.
\label{eq:shapelet}
\end{equation}
This representation does not assume a fixed analytic pulse shape; instead, it provides a flexible basis expansion whose complexity is determined by the data. The expansion order $n_s$ is selected automatically using the corrected Akaike information criterion (AICc), while $(\tau,\beta)$ are optimized via a bounded nonlinear optimizer. The TOA is defined as the peak of the reconstructed, baseline-subtracted profile with sub-bin quadratic interpolation.
Although $(\tau,\beta)$ are fitting parameters, the method is classified as non-parametric because the basis representation can approximate arbitrary smooth profiles given sufficient expansion order. This makes the approach particularly useful for complex or overlapping pulse structures where fixed Gaussian decomposition becomes degenerate (Section~\ref{sec:results}).

\subsection{Sub-Band Cross-Validation} \label{sec:subband}

For pulses in which overlapping components lead to unstable parametric fits, we implement a sub-band cross-validation approach. The dedispersed dynamic spectrum is divided into $N_{\rm sub}$ frequency sub-bands of equal bandwidth. A parametric model (default: single Gaussian) is then fitted independently to the pulse profile in each sub-band.
Because the data are already dedispersed to the top of the band, the resulting TOAs are expected to be consistent within statistical uncertainties. Deviations between sub-bands indicate either residual dispersion measure errors, frequency-dependent pulse morphology, or unresolved substructure.

We quantify the cross-band consistency using the scatter of sub-band TOAs,
\begin{equation}
\sigma_{\rm sub} = \frac{\mathrm{std}(\mu_i)}{\sqrt{N_{\rm valid}}},
\label{eq:sub_scatter}
\end{equation}
where $N_{\rm valid}$ is the number of sub-bands with S/N $\geq 3$. This provides an empirical estimate of the uncertainty associated with frequency-dependent systematics.
Robust averaging is performed after removing low-S/N sub-bands and rejecting outliers using a median-based criterion prior to computing the final TOA.

\subsection{Frequency Range Selection} \label{sec:freq_range}

The 1D pulse profile is formed by summing the dedispersed dynamic spectrum over frequency. By default, the full observing bandwidth is used. However, the analysis band can be restricted to a sub-range in order to exclude low-quality band edges or to isolate specific spectral regions of interest.
Dedispersion and RFI excision are performed on the full band prior to any frequency selection, ensuring that frequency selection does not affect the mitigation of radio-frequency interference. The frequency-restricted profile is then formed by summing only the selected channels.

This frequency selection is complementary to sub-band cross-validation (Section~\ref{sec:subband}), which instead partitions a fixed analysis band to assess frequency-dependent systematics. In contrast, frequency selection defines the global analysis window, while sub-band analysis probes internal consistency within that window.
The reference frequency associated with each TOA is defined independently of the selected band, and can be chosen to represent the top, center, or a user-specified value, allowing consistent comparison across different analysis choices.

\subsection{Time-Resolution Optimization} \label{sec:auto_res}

The S/N of the one-dimensional pulse profile depends on the sampling of the intrinsic pulse width. When the profile is oversampled, rebinning can improve the effective S/N by reducing noise per bin without significantly affecting the pulse peak. In contrast, for well-sampled or narrow pulses, the native resolution is already close to optimal.

We implement an automatic resolution selection procedure (\texttt{--auto\_res}) that evaluates a discrete set of dyadic rebinning factors $f = 2^k$, with $k = 0, 1, \dots, 8$, and computes the peak S/N of the resulting 1D profile. The factor that maximizes the S/N is selected as the optimal resolution. A minimum sampling criterion is imposed to ensure that the pulse remains sufficiently resolved across the fitting window.

The resulting optimally binned profile is then used for subsequent TOA extraction. The signal-to-noise improvement,
\begin{equation}
\mathrm{S/N~gain} \equiv \frac{\mathrm{S/N}(f_{\rm opt})}{\mathrm{S/N}(f=1)},
\end{equation}
is reported as a diagnostic of the degree of oversampling in the original data.

\subsection{Exponentially-Modified Gaussian and Voigt Models} \label{sec:emg_voigt}

Highly asymmetric pulses with scattering tails, common in FRBs and in some search-mode observations (e.g., FAST), are often not well described by symmetric Gaussian profiles. We therefore implement two additional profile models to capture asymmetric pulse shapes.

{\bf EMG.}
The exponentially-modified Gaussian is defined as the convolution of a Gaussian $G(t; \mu, \sigma)$ with an exponential decay $\exp(-t/\tau)\Theta(t)$, producing a profile with a sharp leading edge and an exponential tail. In this model, $\mu$ represents the intrinsic (unscattered) arrival time and $\tau$ characterizes the scattering timescale.

{\bf Voigt.}
The Voigt profile is the convolution of a Gaussian with a Lorentzian of half-width at half-maximum $\gamma$, capturing combined broadening effects. It is evaluated using the standard Faddeeva function.

Both models are available through the \texttt{--profile\_model} option and are fully compatible with all fitting strategies, AICc-based model selection, and MCMC sampling.
In this work, the main RRAT analysis is performed using Gaussian models for consistency, while EMG and Voigt models are provided for general applicability and for future analysis of strongly asymmetric or scattering-dominated pulses.

\subsection{MCMC Uncertainty Estimation} \label{sec:mcmc}

The formal uncertainties returned by \texttt{curve\_fit} rely on a local linear approximation to the likelihood and may be affected by parameter degeneracies in multi-component models (e.g., Section~\ref{sec:results}). To obtain more robust uncertainty estimates, we optionally perform Markov Chain Monte Carlo (MCMC) sampling using \texttt{emcee} \citep{emcee13}.
Walkers are initialized in the vicinity of the best-fit solution, with the covariance matrix providing the initial scale. We adopt flat priors with physically motivated bounds, requiring $\sigma > 0$ and restricting component centroids $\mu_i$ to lie within the fitting window. Convergence is assessed using the Gelman--Rubin $\hat{R}$ statistic, with values close to unity indicating well-mixed chains.
The TOA uncertainty is defined as the posterior standard deviation of the centroid parameter $\mu$, and we additionally report the 16th--84th percentile credible interval.

\subsection{Model Selection} \label{sec:aicc}

The corrected Akaike Information Criterion (AICc) \citep{Akaike74} is used to select the optimal number of Gaussian components:
\begin{equation}
{\rm AICc} = N_{\rm data} \ln({\rm RSS}/N_{\rm data}) + 2k +
\frac{2k(k+1)}{N_{\rm data} - k - 1},
\label{eq:aicc}
\end{equation}
where $N_{\rm data}$ is the number of profile bins in the fit window, $k = 3N+1$ is the number of free parameters, and RSS is the residual sum of squares. AICc penalizes model complexity and is preferred over AIC in the finite-sample regime where $N_{\rm data}/k \lesssim 40$.
Because AICc includes a log-likelihood term that scales with the data amplitude, its absolute values depend on the signal-to-noise ratio of the profile. However, model selection is based on differences in AICc between competing models fitted to the same data, which remain well-defined and invariant under such scaling.

\subsection{Implementation} \label{sec:implementation}

\toacode{} is implemented in Python 3 and relies on standard scientific libraries including \textsc{numpy}, \textsc{scipy}, \textsc{astropy}, and \textsc{matplotlib}. Optional functionality includes MCMC sampling via \texttt{emcee} \citep{emcee13} and posterior visualization using \texttt{corner}.
We benchmark the full pipeline on a 688-pulse RRAT dataset from a single 3-hour FAST observation, consisting of 1706 PSRFITS-format sub-integrations, full polarization data, 4096 frequency channels, 8-bit sampling, and a time resolution of 49.152\,$\mu$s. End-to-end processing, including data loading, dedispersion, RFI excision, profile construction, and evaluation of all TOA strategies, requires $\sim$7.6\,s per pulse on a single CPU using 10 threads.

For sources with sparse single-pulse activity, such computational cost is negligible compared to typical observing durations. MCMC sampling increases runtime by approximately 30--60\,s per pulse and is used only when uncertainty estimates from local optimization are insufficient.

The pipeline produces standardized outputs including TOAs, uncertainties, and metadata such as reference frequency and observatory identifiers. Output files are compatible with TEMPO2 \citep{Hobbs06,Edwards06}. 
The package is available via \texttt{pip install toa\_sp} and the source code is hosted at \url{https://github.com/songqiii/toa\_sp}.

\section{Sample} \label{sec:sample}

\subsection{Source Selection}

We select two source classes spanning the profile complexity spectrum introduced in Section~\ref{sec:intro}. As a definitive test of extreme pulse-to-pulse variability, we use RRAT~J1913$+$1330. Individual pulses from this source span widths of 0.15--17.29\,ms and fluences of 0.005--5.115\,Jy\,ms \citep{Zhang24}, making it an ideal testbed for direct single-pulse timing in the extreme-nulling regime. To complement this, we perform a controlled multi-component test by applying \toacode{} to a subset of bright bursts from FRB~20220529 \citep{Li26}. These bursts span a range of signal-to-noise ratios and sub-component complexity, enabling a rigorous multi-strategy comparison.

\subsection{Observations and Data Reduction}

All data were recorded with the 19-beam receiver of FAST \citep{Jiang20}, covering 1000--1500\,MHz with 4096 frequency channels. Signals were digitized by the ROACH\,2 backend \citep{Hickish16} at 8-bit resolution with a sampling time of 49.152\,$\mu$s and stored in PSRFITS search-mode format \citep{Hotan04}. A 10\,K equivalent noise-diode calibration signal was recorded before each observation.

For RRAT~J1913$+$1330, we use a 3-hour observation on 2019 December 16 (MJD~58833), which is the longest single observation from a set of five FAST epochs reported by \citet{Zhang24}. We adopt their fixed DM of 175.25\,pc\,cm$^{-3}$ \citep{Zhang24}.

For FRB~20220529, we present the results of three bright bursts (M01\_0096, M01\_0322, M01\_0354) from \citet{Li26}, with DMs of 249.95, 244.07, and 255.90\,pc\,cm$^{-3}$, respectively. These bursts span a range of signal-to-noise ratios and sub-component complexity, enabling a controlled multi-strategy comparison.

All data are processed with per-channel normalization to a fixed noise level in arbitrary units and combined into Stokes I by summing the two linear polarization channels. Throughout this work, S/N is defined as the ratio of the baseline-subtracted peak amplitude to the off-pulse RMS measured after scaling and binning. This normalization affects absolute flux units but preserves all relative timing comparisons between methods.

For the PSRCHIVE comparison (Section~\ref{sec:results}), the same search-mode data are summed across all sub-integrations, then processed using \textsc{dspsr} \citep{vanStraten11} and \textsc{psrchive} \citep{vanStraten12} following the standard pipeline, including RFI excision, scrunching in frequency and time, and matched-filter TOA extraction. Both pipelines use the same DM (175.25\,pc\,cm$^{-3}$), ephemeris, and timing model; differences arise solely from the TOA extraction methodology.

\subsection{TOA Extraction Setup}

For RRAT~J1913$+$1330, all 688 pulses are processed using the full comparison pipeline, which evaluates all TOA strategies and selects the final result using the AICc-based model selection combined with the convergence criterion $\Delta_{\rm conv}$ (Section~\ref{sec:conv_tool}). The leading-edge estimator is also evaluated independently as a representative non-parametric baseline.

For FRB~20220529, we apply the full multi-strategy framework to each burst, exploring Gaussian models with component numbers $N=1$--$6$ and enabling sub-band cross-validation. The number of sub-bands is adjusted between bursts to ensure adequate S/N per sub-band while preserving frequency resolution, with higher resolution used for higher-S/N events.

All binning and windowing parameters are chosen to ensure that each pulse is fully contained within the analysis window and sufficiently sampled for all tested rebinning factors.

\section{Results} \label{sec:results}

\subsection{Direct Single-Pulse Timing of RRAT J1913+1330}

\begin{figure*}
\centering
\includegraphics[width=\textwidth]{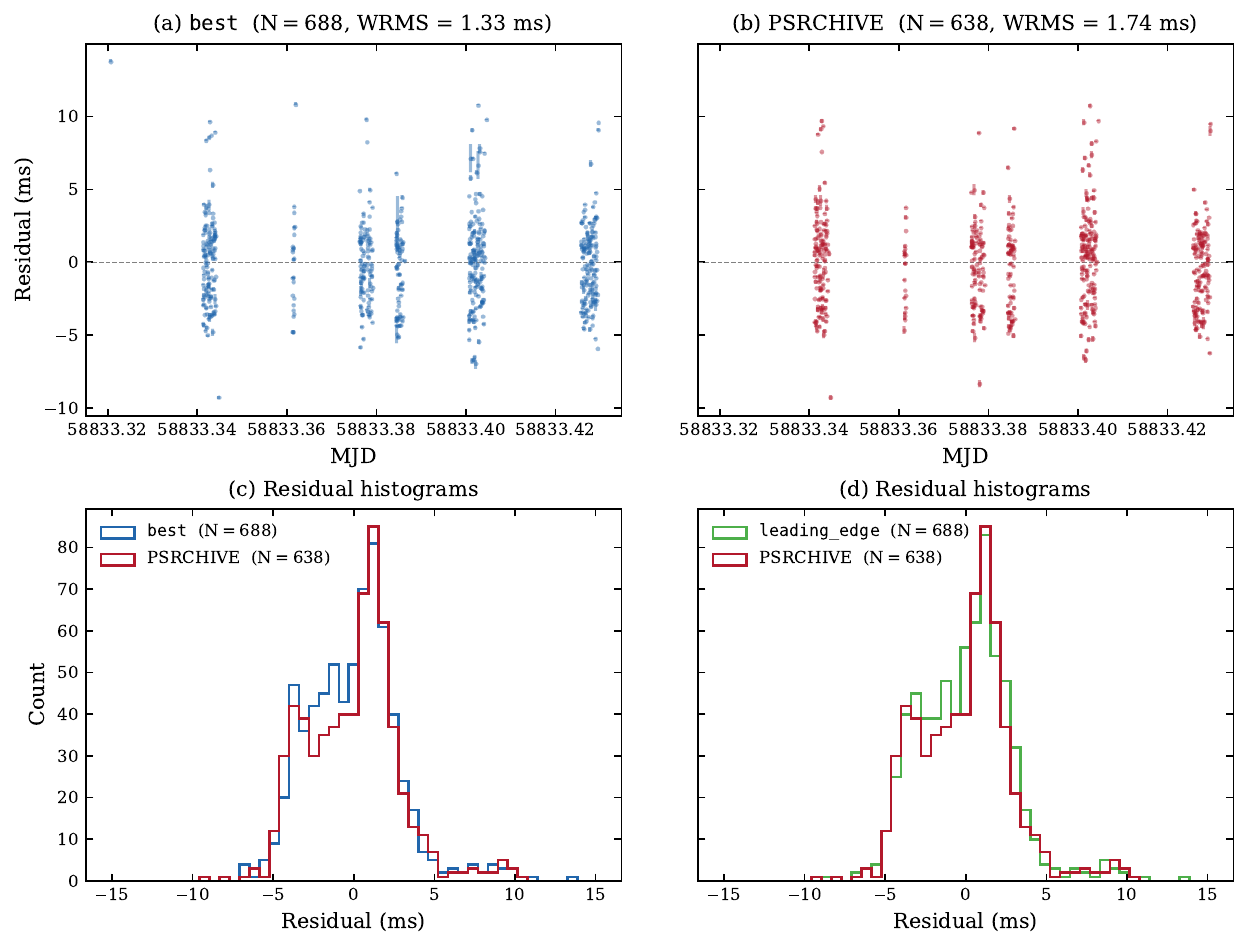}
\caption{Timing residual comparison for RRAT~J1913$+$1330. {\em (a)} Residuals from the \texttt{best} strategy ($N=688$). Weighted RMS = 1.33\,ms. {\em (b)} \textsc{psrchive} results ($N=638$ after standard outlier rejection). Weighted RMS = 1.74\,ms. {\em (c)} Residual histograms. {\em (d)} Leading-edge estimator: weighted RMS = 1.41\,ms.}
\label{fig:rrat_residuals}
\end{figure*}

We apply \toacode{} to the 688 single pulses of RRAT~J1913$+$1330, extracting barycentric arrival times using TEMPO2 \citep{Hobbs06,Edwards06}. Figure~\ref{fig:rrat_residuals} compares three representative strategies: \texttt{best} (AICc $+$ $\Delta_{\rm conv}$ selection), \texttt{leading\_edge} (non-parametric estimator), and the standard \textsc{psrchive}-based folding pipeline \citep{vanStraten12}. 

The full comparison is summarized in Table~\ref{tab:rrat_all}. The \texttt{best} strategy achieves the lowest weighted RMS residual (1.33\,ms), followed closely by the non-parametric \texttt{peak} (1.37\,ms) and \texttt{leading\_edge} (1.41\,ms) estimators. The \texttt{center\_of\_mass} method yields slightly larger scatter (1.61\,ms) and fails for two low-S/N pulses due to zero TOA uncertainty estimates. The \texttt{shapelet} approach performs less well in this dataset (2.18\,ms), likely reflecting limited constraints on basis complexity in the moderate S/N regime of RRAT single pulses.

\begin{table}
\caption{Weighted RMS residuals for RRAT~J1913$+$1330 across all evaluated strategies. All \toacode{} methods process the full set of 688 single pulses; the \textsc{psrchive} pipeline reports 638 TOAs after standard quality filtering.}
\label{tab:rrat_all}
\begin{tabular}{lcc}
\hline
Strategy & $N_{\rm TOA}$ & Weighted RMS (ms) \\
\hline
\texttt{best} (AICc $+$ $\Delta_{\rm conv}$) & 688 & 1.33 \\
\texttt{peak}           & 688 & 1.37 \\
\texttt{leading\_edge}  & 688 & 1.41 \\
\texttt{center\_of\_mass}$^\dagger$ & 686 & 1.61 \\
\textsc{psrchive} (folding) & 638 & 1.74 \\
\texttt{shapelet}       & 688 & 2.18 \\
\hline
\end{tabular}
$^\dagger$ Two pulses excluded due to ill-conditioned TOA uncertainty estimates.
\end{table}

Our method retains all 688 pulses without outlier rejection, yielding a weighted RMS residual of 1.33\,ms, compared to 1.74\,ms for the \textsc{psrchive} pipeline after standard outlier removal. The difference reflects the treatment of highly variable pulses whose morphologies deviate from the folded template, which can increase scatter in template-based timing when no additional pulse-level selection is applied. In contrast, direct single-pulse timing avoids reliance on a fixed integrated profile and treats each pulse individually.

\subsection{Multi-Strategy Tests on FRB 20220529: The Role of Sub-Band Structure}

Having demonstrated that direct single-pulse timing performs well for highly variable RRATs, we now apply \toacode{} to FRB~20220529 as a controlled test of the complete multi-strategy framework. Compared with RRATs, FRB bursts often exhibit pronounced frequency-dependent pulse morphology: the number of sub-components, their relative amplitudes, and even their apparent arrival order can vary across the observing band. Consequently, TOAs derived from a single band-integrated profile may not fully represent the underlying burst structure.

To investigate these effects, we evaluate all nine TOA estimation strategies, together with sub-band cross-validation, on three representative bursts spanning a wide range of signal-to-noise ratios (S/N = 8--271) and profile complexity.

\subsubsection{Case 1: Well-Separated Components (M01\_0096)}

\begin{figure}
\centering
\includegraphics[width=\columnwidth]{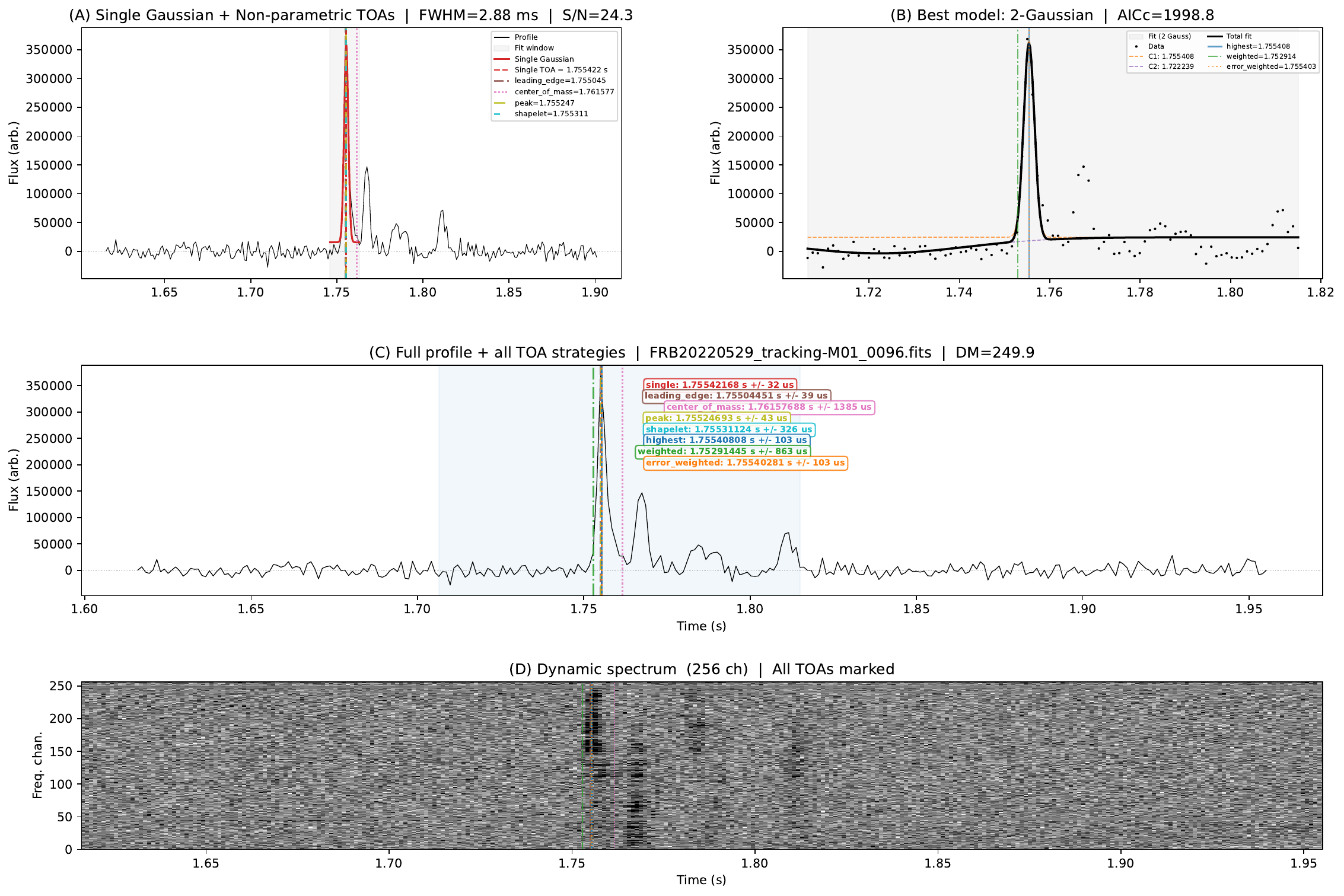}
\caption{TOA strategy comparison for FRB~20220529 burst M01\_0096 (S/N = 24.3) containing four well-separated emission components. {\em Top left:} Single-Gaussian fit with non-parametric TOA estimates. {\em Top right:} Best multi-Gaussian fit ($N=2$) with all parametric TOAs. {\em Bottom left:} Full pulse profile with all TOA estimates annotated. {\em Bottom right:} Dynamic spectrum with corresponding TOA overlays. The robust TOA estimators agree within $\pm14\,\mu$s, well inside the empirical uncertainty of $3\sigma_{\rm emp}=95\,\mu$s.}
\label{fig:case1}
\end{figure}

Burst M01\_0096 contains four significant emission components, with the brightest reaching S/N = 24.3. The components are separated by 12--57\,ms, substantially larger than their typical FWHM of $\sim$3\,ms, making this burst representative of a regime in which individual components are well resolved. Figure~\ref{fig:case1} compares the TOAs obtained using different strategies.

The \texttt{single}, \texttt{highest} ($N=2$), \texttt{error\_weighted}, and \texttt{first\_peak} strategies produce TOAs that agree within $\pm22\,\mu$s, smaller than the empirical uncertainty of $\sigma_{\rm emp}=31.5\,\mu$s. AICc strongly favors the single-Gaussian model over multi-component alternatives, indicating that introducing additional Gaussian components is not justified by the improvement in fit quality.

The only exception is the deprecated \texttt{weighted} strategy, which yields a TOA offset of 2.5\,ms. In this case, a low-S/N component with negative amplitude receives substantial weight in the amplitude-weighted average, illustrating the sensitivity of this strategy to poorly constrained fitted components.

Sub-band cross-validation ($N_{\rm sub}=8$) supports this interpretation. All eight sub-bands produce mutually consistent TOAs within approximately $\pm30\,\mu$s, giving $\sigma_{\rm sub} \ll \sigma_{\rm emp}$, which indicates that frequency-dependent profile evolution is negligible for this burst.

To assess whether the formal covariance matrix provides realistic parameter uncertainties, we additionally performed MCMC sampling of the single-Gaussian model. The chains converged well ($\hat{R}=1.057$), yielding a posterior TOA uncertainty of 53.6\,$\mu$s, compared with the formal covariance estimate of 31.5\,$\mu$s. Although the formal uncertainty is underestimated by a factor of $\sim1.7$, both estimates remain much smaller than the observed agreement among the different TOA strategies, indicating that the pulse morphology is sufficiently simple for all robust estimators to converge.

Overall, this burst represents the regime in which direct single-pulse timing is straightforward. When the emission components are well separated, the choice among the recommended parametric strategies has little impact on the inferred TOA, while non-parametric estimates provide consistent independent validation. Only the deprecated amplitude-weighted strategy exhibits pathological behavior and should therefore be avoided.

\subsubsection{Case 2: Densely Overlapping Components (M01\_0322)}

\begin{figure}
\centering
\includegraphics[width=\columnwidth]{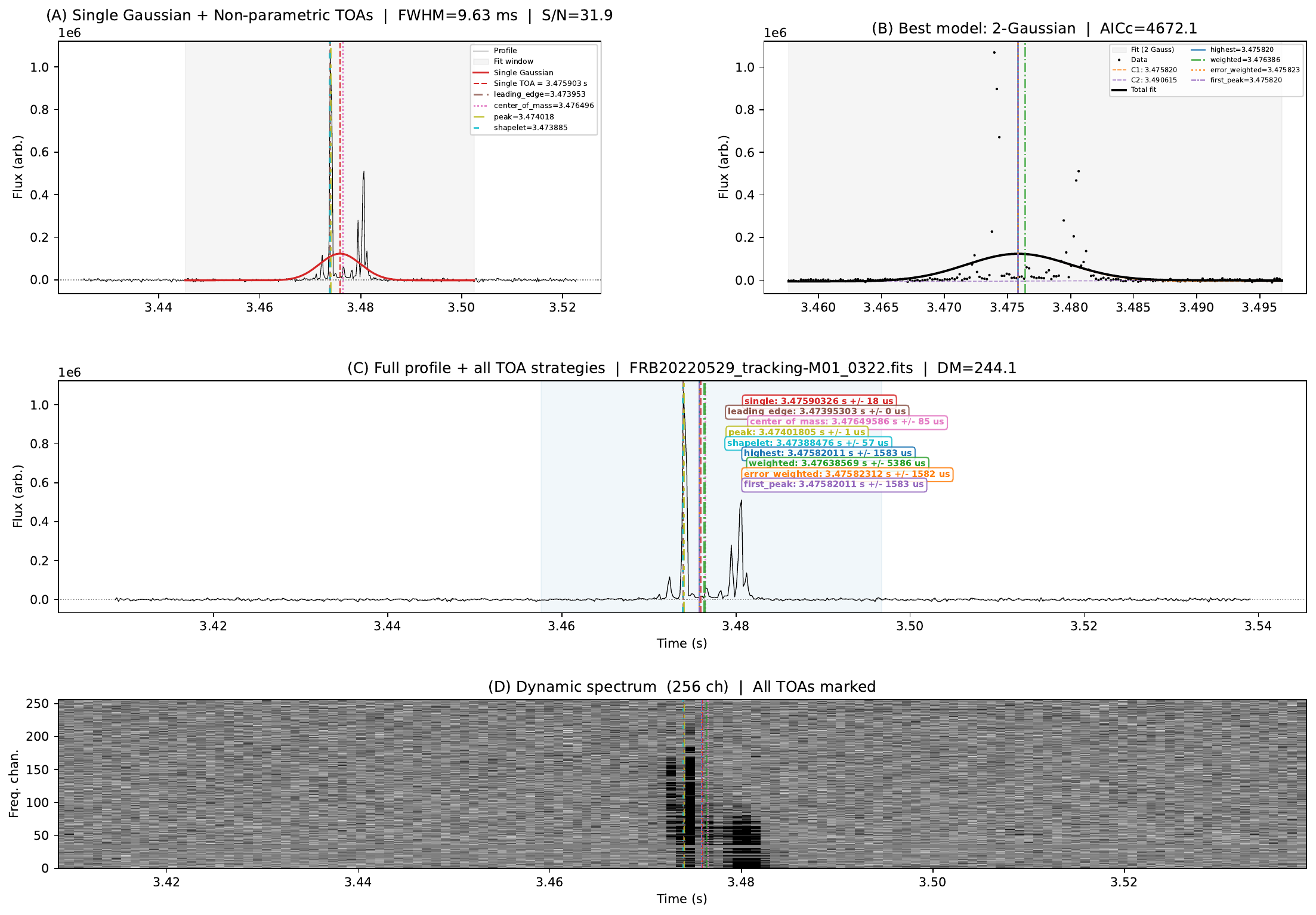}
\caption{TOA strategy comparison for FRB~20220529 burst M01\_0322 (dense overlapping regime). {\em Top left:} Single-Gaussian fit with non-parametric TOAs. {\em Top right:} Best multi-Gaussian representation ($N=2$). {\em Bottom left:} Full profile with all strategies overplotted, showing multiple TOA clusters separated by several milliseconds. {\em Bottom right:} Dynamic spectrum. The parametric \texttt{highest} estimator exhibits discontinuous shifts of order several milliseconds as $N$ changes, indicating strong sensitivity to decomposition choice.}
\label{fig:case2}
\end{figure}

Burst M01\_0322 exhibits a highly structured morphology, consisting of multiple partially overlapping emission components distributed over a $\sim$10\,ms envelope. The typical separation between adjacent peaks ($\sim$1\,ms) is comparable to the intrinsic component widths, placing this burst in a regime where individual components are not cleanly resolved.
Figure~\ref{fig:case2} compares the behavior of different TOA strategies in this complex regime.

In this regime, the \texttt{highest} strategy becomes unstable, with TOA shifts of up to 8.3\,ms depending on the number of fitted components. This behavior arises from the intrinsic degeneracy of Gaussian decomposition when multiple overlapping components redistribute flux among neighboring fits. As $N$ increases, the identity of the brightest component is no longer invariant, leading to discontinuous changes in the inferred TOA.
Although AICc mildly favors a two-component model, higher-order decompositions yield statistically similar fits, and begin to capture noise fluctuations as spurious narrow components. This indicates that model selection alone is insufficient to regularize the decomposition in the presence of strong overlap.

Sub-band cross-validation ($N_{\rm sub}=8$) reveals moderate but non-negligible frequency dependence, with a scatter of $\sigma_{\rm sub} \approx 0.5$\,ms. This scatter is significantly smaller than the inter-model variability but larger than the formal statistical uncertainties, indicating the presence of unresolved structure rather than pure noise.
MCMC sampling of the single-Gaussian model ($\hat{R}=1.047$) increases the TOA uncertainty to 32.4\,$\mu$s, a factor of $\sim1.8$ above the covariance estimate. This confirms that linear error propagation becomes unreliable in the presence of overlapping structure, although the dominant source of uncertainty is still model ambiguity rather than sampling noise.

Overall, this burst marks the transition regime where parametric decomposition becomes non-unique. In this case, stable TOAs are obtained only from estimators that do not depend on component identification, such as \texttt{single} and \texttt{leading\_edge}.

\subsubsection{Case 3:  Broad Low-S/N Envelope with Frequency Evolution (M01\_0354)}

Burst M01\_0354 represents the low-S/N regime (S/N = 8), characterized by a broad envelope (FWHM $\sim$42\,ms) and weak sub-structure. Unlike the previous cases, the dominant source of uncertainty here is not component overlap but frequency-dependent profile evolution.

\begin{figure}
\centering
\includegraphics[width=\columnwidth]{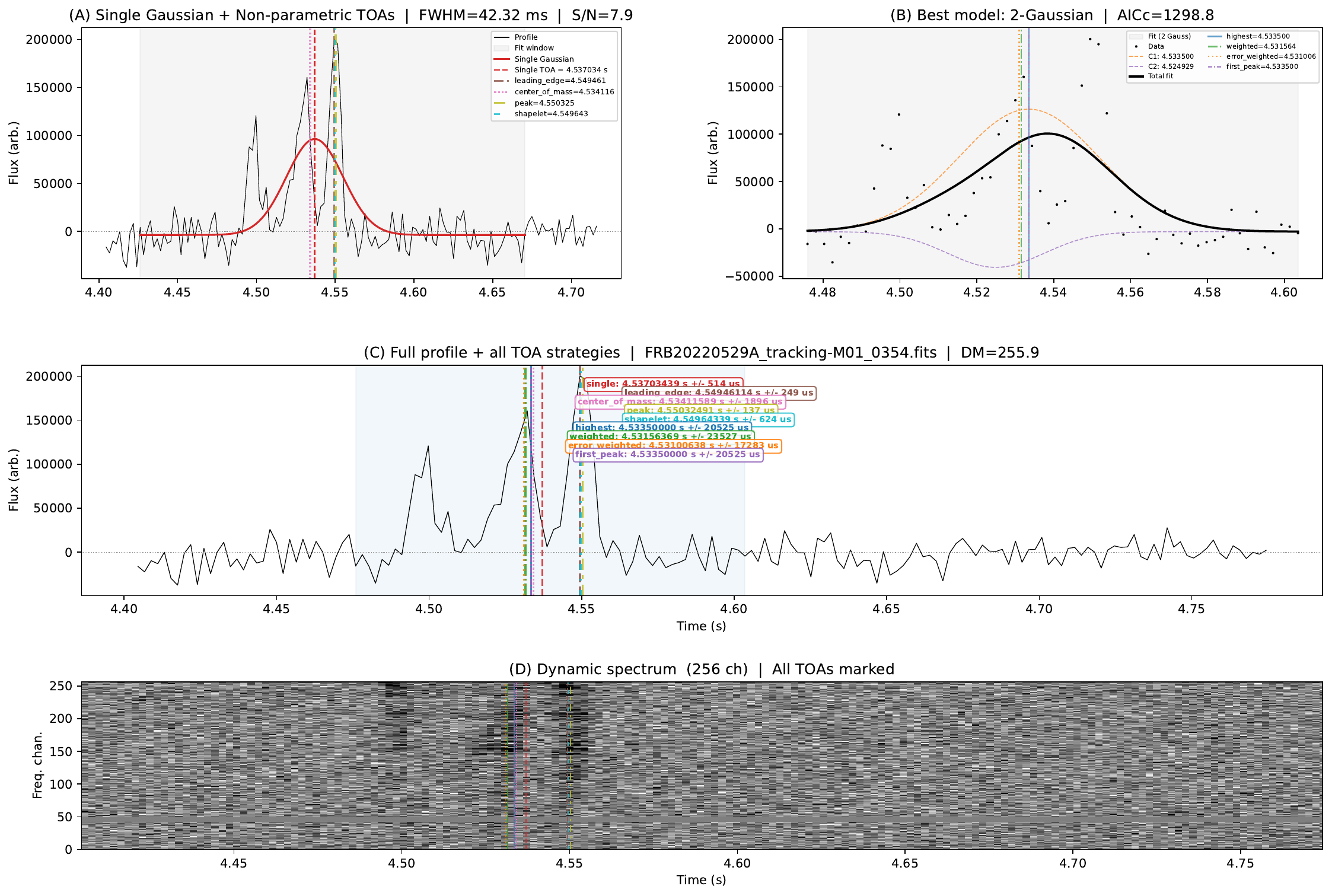}
\caption{Sub-band TOA analysis for FRB~20220529 burst M01\_0354 (low S/N, broad envelope). {\em Left:} Multi-strategy TOA comparison showing a total spread of 16.5\,ms across estimators. {\em Right:} Sub-band TOAs measured in three frequency bands after S/N thresholding. A systematic trend of increasing TOA with frequency is observed, with a total shift of $\sim$18\,ms across the band.}
\label{fig:case3}
\end{figure}

Using $N_{\rm sub}=4$, three frequency sub-bands satisfy the S/N $>3$ criterion: 1.00--1.12\,GHz, 1.25--1.37\,GHz, and 1.37--1.50\,GHz. The corresponding TOAs exhibit a monotonic increase with frequency, producing a total drift of $\sim$18\,ms across the band. This variation is orders of magnitude larger than the expected residual dispersion error ($\sim0.02$\,ms for the given bandwidth and DM), indicating that it cannot be attributed to imperfect dedispersion.
Instead, the observed trend is consistent with frequency-dependent pulse morphology, likely arising from unresolved sub-components whose relative contributions evolve across the band. In this regime, sub-band cross-validation yields a scatter of $\sigma_{\rm sub}=4.3$\,ms, which significantly exceeds both the formal statistical uncertainties and the differences among most parametric estimators.

MCMC sampling further increases the TOA uncertainty to 1.06\,ms, approximately twice the covariance-based estimate, but still substantially below the systematic scatter measured across frequency. This demonstrates that, for low-S/N bursts with strong spectral evolution, statistical and sampling uncertainties are subdominant compared to frequency-dependent systematics.
In this regime, sub-band analysis provides the most reliable estimate of TOA uncertainty, while single-profile fitting is insufficient to capture the full error budget.

\subsubsection{Summary of FRB Regimes}

The three FRB cases together define a hierarchy of increasingly complex timing regimes.

In the well-separated regime (M01\_0096), all robust estimators converge to a consistent TOA, and the primary uncertainty is purely statistical. In this case, both parametric and non-parametric methods are interchangeable.

In the overlapping-component regime (M01\_0322), the dominant limitation becomes the non-uniqueness of parametric decomposition. Different choices of Gaussian component number $N$ redistribute flux among fitted components, leading to discontinuous shifts in the inferred TOA. Here, estimators that do not rely on component identification (e.g.\ \texttt{single} and \texttt{leading\_edge}) remain stable, while decomposition-dependent methods become unreliable.

In the low-S/N, frequency-evolving regime (M01\_0354), the dominant source of uncertainty is no longer model ambiguity but frequency-dependent pulse morphology. Sub-band cross-validation reveals systematic TOA drifts across the observing band that exceed both statistical and sampling uncertainties, indicating that the effective TOA is intrinsically frequency dependent.

Together, these regimes demonstrate that single-pulse timing is governed by three distinct error classes: statistical uncertainty, model degeneracy, and frequency-dependent systematics. The quantities $\Delta_{\rm conv}$ and $\sigma_{\rm sub}$ provide complementary diagnostics for identifying the first two regimes and diagnosing the transition into the third. This framework enables a unified interpretation of TOA reliability across both RRATs and FRBs.

\subsection{Strategy Reliability Matrix and the Convergence Diagnostic}

Table~\ref{tab:cases} summarizes the diagnostic behavior of the three FRB case studies across all TOA strategies, while Table~\ref{tab:reliability} synthesizes their performance across different complexity regimes, including an additional control configuration for M01\_0354 with finer time binning (bs=10).

\begin{table}
\caption{Summary of the three FRB 20220529 case studies.}
\label{tab:cases}
\centering
\begin{tabular}{lcccc}
\hline
Burst & S/N & $\Delta_{\rm conv}$ & $3\sigma_{\rm emp}$ & Recommended \\
\hline
M01\_0096 & 24.3 & 14\,$\mu$s & 95\,$\mu$s & single (AICc best) \\
M01\_0322 & 31.9 & 83\,$\mu$s & 55\,$\mu$s & leading\_edge \\
M01\_0354 & 7.9  & 3.5\,ms & 1.5\,ms & leading\_edge \\
\hline
\end{tabular}
\end{table}

\begin{table}
\footnotesize
\caption{TOA strategy reliability by profile regime.  $\checkmark$ = reliable ($\Delta{\rm TOA} < 3\sigma_{\rm emp}$); $\times$ = unreliable; $\triangle$ = reliable but measures a different physical quantity.}
\centering
\label{tab:reliability}
\begin{tabular}{lccc}
\hline
Strategy & Well-separated & Overlapping & Broad, low-S/N \\
\hline
\texttt{single}           & $\checkmark$ & $\checkmark$ & $\checkmark$ \\
\texttt{highest} ($N=2$)  & $\checkmark$ & $\checkmark$ & $\checkmark$ \\
\texttt{highest} ($N\geq3$)& $\checkmark$ & $\times$ & $\times$ \\
\texttt{weighted}         & $\times$ & $\times$ & $\times$ \\
\texttt{error\_weighted}  & $\checkmark$ & $\triangle$ & $\times$ \\
\texttt{first\_peak}      & $\checkmark$ & $\triangle$ & $\triangle$ \\
\texttt{leading\_edge}    & $\triangle$ & $\triangle$ & $\triangle$ \\
\texttt{center\_of\_mass} & $\triangle$ & $\triangle$ & $\triangle$ \\
\texttt{peak}             & $\times$ & $\times$ & $\times$ \\
\texttt{shapelet}         & $\checkmark$ & $\checkmark$ & $\triangle$ \\
Sub-band                 & $\checkmark$ & $\checkmark$ & $\checkmark$ \\
\hline
\end{tabular}
\end{table}

To provide a unified criterion for identifying when parametric strategies remain stable, we define a practical convergence diagnostic based on the sensitivity of Gaussian decomposition:

\begin{equation}
\Delta_{\rm conv} \equiv 
\left| {\rm TOA}(N{=}2,\,{\rm highest}) - {\rm TOA}(N{=}1) \right|.
\label{eq:conv}
\end{equation}

Here, ${\rm TOA}(N{=}1)$ is the single-Gaussian solution, and ${\rm TOA}(N{=}2,{\rm highest})$ is the brightest-component solution from a two-component fit. The diagnostic measures the stability of the inferred arrival time under minimal model complexity increase.

We compare $\Delta_{\rm conv}$ to the empirical single-Gaussian uncertainty,
\begin{equation}
\sigma_{\rm emp} \approx \frac{\sigma}{\mathrm{S/N}\sqrt{N_{\rm eff}}},
\end{equation}
and classify pulses as follows:

{\bf Well-behaved regime:} $\Delta_{\rm conv} < 3\sigma_{\rm emp}$, where parametric TOAs are stable and either \texttt{single} or \texttt{highest} ($N=2$) provides reliable results.

{\bf Overlapping regime:} $\Delta_{\rm conv} \gtrsim 3\sigma_{\rm emp}$, where Gaussian decomposition becomes non-unique and parametric strategies diverge; robust estimates require either \texttt{single} or non-parametric/sub-band approaches.

For the three FRB bursts:

M01\_0096 has $\Delta_{\rm conv} = 14\,\mu$s, which is well below $3\sigma_{\rm emp} \approx 95\,\mu$s, indicating a stable decomposition regime.

M01\_0322 shows $\Delta_{\rm conv} = 83\,\mu$s, exceeding $3\sigma_{\rm emp} \approx 55\,\mu$s. This places it in the transition regime where component assignments become non-unique, although several strategies still agree within $\sim 0.1$\,ms.

M01\_0354 exhibits $\Delta_{\rm conv} = 3.5$\,ms, significantly larger than $3\sigma_{\rm emp} \approx 1.5$\,ms, indicating a strongly overlapping, low-S/N regime in which sub-band cross-validation provides the most reliable uncertainty estimate.

Together, $\Delta_{\rm conv}$ quantifies decomposition stability, while sub-band scatter $\sigma_{\rm sub}$ captures frequency-dependent structure; the two diagnostics are complementary and jointly define the reliability of single-pulse TOA measurements.

\section{Discussion} \label{sec:discussion}

\subsection{Why Amplitude-Weighted Averaging Fails}

The \texttt{weighted} strategy implicitly assumes that Gaussian components provide a meaningful decomposition of the pulse into physically interpretable sub-structures. In practice, this assumption breaks down for overlapping or under-constrained fits. Components with artificially large amplitudes can arise when the fitting procedure assigns excess flux to broadened Gaussians that partially absorb noise fluctuations or baseline structure.

Because the weighted TOA scales directly with $|A_i|$, such components can disproportionately influence the final estimate even when their centroid locations are poorly constrained. This effect is particularly pronounced in multi-component fits ($N \geq 3$), where negative-amplitude components may also appear as part of the degeneracy between overlapping structures and baseline absorption terms.

The resulting bias is systematic rather than stochastic, arising from the non-uniqueness of Gaussian decompositions rather than from statistical noise. This explains the instability observed in Section~\ref{sec:results}, and suggests that amplitude-weighted averaging is not robust for general single-pulse timing in complex profiles.

\subsection{The Convergence Diagnostic as a Practical Tool}
\label{sec:conv_tool}

The $\Delta_{\rm conv}$ diagnostic (Eq.~\ref{eq:conv}) provides a simple measure of the stability of Gaussian decomposition under a minimal increase in model complexity. It is implemented in \toacode{} as part of the automated strategy selection procedure (\texttt{--tim\_strategy best}).

{\bf Stability regime ($\Delta_{\rm conv} < 3\sigma_{\rm emp}$).}
In this regime, the TOA is relatively insensitive to the choice between a single-component model and a minimal multi-component extension. Parametric strategies are typically sufficient, and differences between \texttt{single} and \texttt{highest} ($N=2$) remain within the expected statistical uncertainty. In this case, AICc provides an additional tie-breaker for model selection.

{\bf Unstable regime ($\Delta_{\rm conv} \gtrsim 3\sigma_{\rm emp}$).}
Here the inferred TOA becomes sensitive to how flux is partitioned among overlapping components, indicating that the Gaussian decomposition is not unique. In such cases, non-parametric approaches provide more robust alternatives by reducing dependence on component identities.

Among these, the \texttt{leading\_edge} estimator is generally the most stable because it relies on the rising portion of the pulse, which is less affected by scattering-induced asymmetries. The \texttt{shapelet} method can also perform well when the signal-to-noise ratio is sufficient for a stable basis expansion, as it effectively denoises the profile before TOA extraction. When neither approach is applicable, the single-Gaussian estimate may still be used, but its limitations should be explicitly reflected in the reported uncertainties.

{\bf Role of \texttt{center\_of\_mass}.}
The flux-weighted centroid is sensitive to profile asymmetry, particularly in the presence of scattering tails or residual instrumental smearing after incoherent dedispersion. In such cases, additional power on the trailing side biases the centroid toward later times without corresponding to a true shift in the intrinsic emission time.
While this effect is minimized for symmetric profiles or fully coherently dedispersed data, it is generally non-negligible for most single-pulse observations. For this reason, the centroid is not used in the automated selection pipeline in \toacode{}, although it remains available for exploratory analysis.

\subsection{Computational Efficiency}

Table~\ref{tab:efficiency} compares the processing time of the \toacode{} pipeline with a traditional \textsc{dspsr} + \textsc{psrchive} workflow for a representative FAST observation (1706 \textsc{PSRFITS}-format sub-integration, 4096 channels, $\sim$three hour).
The conventional pipeline requires first combining all sub-integrations before folding, which introduces a substantial preprocessing overhead in addition to the subsequent TOA extraction. In contrast, \toacode{} operates directly on individual sub-integration files, allowing processing to begin immediately and enabling a streaming-style workflow in which cost scales with the number of detected pulses rather than total observation volume.

\begin{table*}
\caption{Processing time comparison for a typical FAST three-hour observation.}
\label{tab:efficiency}
\centering
\begin{tabular}{lcc}
\hline
Pipeline & CPU cores & Wall time \\
\hline
\textsc{dspsr} + \textsc{psrchive} (folding) & 2 & $\sim$4 days \\
\toacode{} (full dataset, 688 pulses) & 1 (10 threads) & $\sim$87 minutes \\
\toacode{} (per pulse average) & 1 & $\sim$7.6 seconds \\
\hline
\end{tabular}
\end{table*}

The performance difference reflects several structural differences between the two approaches. First, \toacode{} avoids full-stream folding and instead processes only localized pulse windows. Second, all preprocessing steps, including I/O, unpacking, and dedispersion, are implemented in vectorized form. Third, no template construction or iterative cross-correlation is required. Finally, the per-pulse structure allows trivial parallelization across events.
For surveys such as CRAFTS \citep{Li18}, where large volumes of search-mode data must be processed, this difference in scaling behavior can significantly reduce end-to-end processing time, particularly in low-duty-cycle regimes such as RRAT and FRB searches.

\subsection{Sub-band Scatter as an Uncertainty Calibrator}

The sub-band cross-validation method (Section~\ref{sec:subband}) provides an additional diagnostic beyond TOA extraction by probing the consistency of the inferred arrival time across frequency.
In Case 3 (M01\_0354), the empirical uncertainty from single-Gaussian fitting (514\,$\mu$s) would typically be adopted as the TOA error in standard single-profile pipelines. However, the sub-band analysis yields a scatter of 4.3\,ms, indicating substantially larger variability across frequency than implied by the parametric uncertainty estimate.

This discrepancy arises because single-profile fitting assumes that the frequency-integrated pulse is adequately described by a single functional form. Sub-band decomposition relaxes this assumption and instead tests whether the inferred TOA is stable across independent frequency slices, thereby providing sensitivity to unresolved spectral structure or frequency-dependent profile evolution.
In this sense, $\sigma_{\rm sub}$ can be interpreted as a practical upper bound on the TOA uncertainty when strong frequency-dependent morphology is present. In cases where $\sigma_{\rm sub} \gg \sigma_{\rm emp}$ (e.g., by more than a factor of a few), reporting both quantities may be more informative than relying on a single parametric uncertainty estimate.

\subsection{Implications for Gravitationally Lensed Pulsars}

A potential future application of direct single-pulse timing is the detection and characterization of pulsars in the presence of compact gravitational lenses, including stellar- or intermediate-mass black holes. In such systems, strong-field lensing can produce multiple propagation paths for individual pulses, leading to a superposition of time-delayed and magnified sub-components \citep{Kimpson20}. If the lensing configuration evolves on orbital timescales, the observed pulse morphology may vary from pulse to pulse, limiting the applicability of standard folding-based timing approaches that rely on a stable integrated profile.

Within this context, several aspects of the \toacode{} framework may be relevant. Multi-component fitting strategies can naturally represent superposed emission features, and the AICc-based selection criterion (Section~\ref{sec:aicc}) provides a data-driven estimate of the effective number of distinguishable components, although it does not by itself encode physical lensing information. In addition, asymmetric profile models such as the EMG (Section~\ref{sec:emg_voigt}) may be useful for describing cases where differential magnification or propagation effects introduce skewness into otherwise symmetric intrinsic pulses, as expected in strong-field environments \citep{Kimpson20,Xiao24}.

Sub-band cross-validation (Section~\ref{sec:subband}) offers a complementary diagnostic by testing the consistency of inferred TOAs across frequency. While gravitational lensing is fundamentally achromatic, frequency-dependent propagation effects such as scattering or dispersion can introduce additional structure that may complicate the interpretation of multi-component profiles. In this sense, sub-band stability may help distinguish between intrinsic or lensing-induced multiplicity and propagation-induced broadening, although a full separation of these effects would likely require additional physical modeling.

In regions such as the Galactic Center, where scattering is expected to be strong at GHz frequencies \citep{Johnston06}, resolving individual pulses remains challenging. Nevertheless, the ability to operate directly on search-mode data and to avoid full folding may be advantageous in regimes where pulse-to-pulse variability dominates over long-term stability. The computational efficiency of the pipeline (Table~\ref{tab:efficiency}) further supports its applicability to large-scale searches of such environments.

\subsection{On Outlier Rejection}

Traditional pulsar timing pipelines often reject TOAs whose post-fit residuals exceed a fixed threshold (e.g., $3\sigma$), typically within an iterative fitting procedure. This approach is well justified in the context of folded profiles, where each TOA represents an average over many rotations and deviations are most often associated with unmodelled systematics, residual radio-frequency interference, or propagation effects.

For single-pulse timing, however, the interpretation of such deviations is fundamentally different. Each TOA corresponds to an individual emission event rather than an ensemble average. As a result, large residuals may reflect intrinsic pulse-to-pulse variability rather than measurement error. In sources such as RRATs and mode-changing pulsars, pulses that appear as statistical outliers in a timing fit can correspond to rare or extreme emission states, including unusually narrow or broad pulses or complex sub-structure.

In this regime, rejecting TOAs solely on the basis of their residuals with respect to a timing model can bias the inferred distribution of emission properties. A more conservative approach is to retain all TOAs unless there is clear evidence of extraction failure. Such failures may include, for example, incomplete dedispersion within the data window, divergence of the profile fit, or unphysical parameter values (e.g., negative widths in Gaussian components). These cases are typically identifiable from diagnostic outputs and comparison summaries produced by \toacode{}.

Rather than enforcing strict rejection criteria, it may therefore be more appropriate to interpret the full distribution of TOAs as part of the astrophysical signal, with the scatter itself carrying physical information about emission variability and propagation effects.

\subsection{Limitations}

Several limitations of the current framework should be noted. The sub-band cross-validation method requires sufficient signal-to-noise per sub-band (typically S/N $\gtrsim 5$) for stable profile fitting, which limits the usable number of sub-bands for faint bursts. In such cases, the diagnostic power of $\sigma_{\rm sub}$ may be reduced.

The current implementation assumes a fixed dispersion measure. Sources exhibiting significant pulse-to-pulse DM variability would require a joint DM–TOA inference framework rather than independent TOA extraction.

Finally, MCMC-based uncertainty estimation improves robustness in cases with strong parameter degeneracies, but increases computational cost by approximately 30–60 seconds per pulse. For high-S/N and morphologically simple pulses, the gain over covariance-based estimates may be marginal, and its use may be unnecessary in routine processing.

\section{Summary and Conclusions} \label{sec:conclusion}

We have presented \toacode{}, an open-source Python package for extracting single-pulse TOAs directly from PSRFITS search-mode data using nine complementary strategies spanning parametric, non-parametric, and sub-band approaches. Applied to 688 single pulses from a 3-hour FAST observation of RRAT~J1913$+$1330, the framework achieves a weighted RMS residual of 1.33\,ms, representing a 24\% improvement over the traditional \textsc{psrchive} folding pipeline (1.74\,ms), while retaining all pulses without statistical outlier rejection. With the exception of the \texttt{shapelet} method, all \toacode{} strategies outperform the folding-based reference, with the non-parametric methods \texttt{peak} (1.37\,ms) and \texttt{leading\_edge} (1.41\,ms) providing competitive precision without assuming a parametric pulse model. Three bright FRB 20220529 bursts further demonstrate robust performance across regimes ranging from well-separated to densely overlapping components.

Two complementary diagnostics provide practical guidance for strategy selection. The convergence metric
$\Delta_{\rm conv} = |{\rm TOA}(N{=}2,\,{\rm highest}) - {\rm TOA}(N{=}1)|$
distinguishes well-behaved pulses, for which parametric strategies are self-consistent and lowest-AICc models are adequate, from overlapping cases where decomposition becomes unstable and non-parametric or sub-band methods are preferred. In parallel, sub-band cross-validation quantifies frequency-dependent profile evolution that is not captured in band-integrated fits; in low-S/N or highly structured pulses, the resulting scatter $\sigma_{\rm sub}$ can exceed the nominal single-profile uncertainty by an order of magnitude and provides a more realistic error estimate.

We advocate retaining all single-pulse TOAs unless a clear algorithmic failure is identified, such as breakdown in dedispersion alignment, fit divergence, or non-physical model parameters. In contrast to folded-profile timing, individual pulses that appear as outliers often reflect intrinsic emission variability rather than statistical contamination, and removing them can bias the inferred timing behavior.

The framework processes a typical FAST observation in $\sim$87 minutes on a single CPU, representing a speed-up of about two orders of magnitude over traditional folding-based pipelines, and enabling efficient analysis of large single-pulse datasets such as CRAFTS \citep{Li18}.

\toacode{} is publicly available via \texttt{pip install toa\_sp} (\url{https://github.com/songqiii/toa\_sp}) and produces TEMPO2-compatible \texttt{.tim} files for downstream timing analysis.

\section*{Acknowledgments}
We thank the FAST staff for their support of the observations. We acknowledge the use of artificial intelligence tools for language editing and code optimization during the preparation of this manuscript. This work is supported by the National Natural Science Foundation of China (grant No.\ 12503058), the Postdoctoral Fellowship Program of CPSF (grant No.\ GZC20252100), the ACAMAR Postdoctoral Fellowship, the China Postdoctoral Science Foundation (grant No.\ 2025M773201), and the Jiangsu Funding Program for Excellent Postdoctoral Talent.

\bibliography{sp_timing}{}

@BOOK{Lorimer04,
       author = {{Lorimer}, D.~R. and {Kramer}, M.},
        title = "{Handbook of Pulsar Astronomy}",
         year = 2004,
       volume = {4},
       adsurl = {https://ui.adsabs.harvard.edu/abs/2004hpa..book.....L}
}

@ARTICLE{Verbiest16,
       author = {{Verbiest}, J.~P.~W. and {Lentati}, L. and {Hobbs}, G. and {van Haasteren}, R. and {Demorest}, P.~B. and {Janssen}, G.~H. and {Wang}, J.-B. and {Desvignes}, G. and {Caballero}, R.~N. and {Keith}, M.~J. and {Champion}, D.~J. and {Arzoumanian}, Z. and {Babak}, S. and {Bassa}, C.~G. and {Bhat}, N.~D.~R. and {Brazier}, A. and {Brem}, P. and {Burgay}, M. and {Burke-Spolaor}, S. and {Chamberlin}, S.~J. and {Chatterjee}, S. and {Christy}, B. and {Cognard}, I. and {Cordes}, J.~M. and {Dai}, S. and {Dolch}, T. and {Ellis}, J.~A. and {Ferdman}, R.~D. and {Fonseca}, E. and {Gair}, J.~R. and {Garver-Daniels}, N.~E. and {Gentile}, P. and {Gonzalez}, M.~E. and {Graikou}, E. and {Guillemot}, L. and {Hessels}, J.~W.~T. and {Jones}, G. and {Karuppusamy}, R. and {Kerr}, M. and {Kramer}, M. and {Lam}, M.~T. and {Lasky}, P.~D. and {Lassus}, A. and {Lazarus}, P. and {Lazio}, T.~J.~W. and {Lee}, K.~J. and {Levin}, L. and {Liu}, K. and {Lynch}, R.~S. and {Lyne}, A.~G. and {Mckee}, J. and {McLaughlin}, M.~A. and {McWilliams}, S.~T. and {Madison}, D.~R. and {Manchester}, R.~N. and {Mingarelli}, C.~M.~F. and {Nice}, D.~J. and {Os{\l}owski}, S. and {Palliyaguru}, N.~T. and {Pennucci}, T.~T. and {Perera}, B.~B.~P. and {Perrodin}, D. and {Possenti}, A. and {Petiteau}, A. and {Ransom}, S.~M. and {Reardon}, D. and {Rosado}, P.~A. and {Sanidas}, S.~A. and {Sesana}, A. and {Shaifullah}, G. and {Shannon}, R.~M. and {Siemens}, X. and {Simon}, J. and {Smits}, R. and {Spiewak}, R. and {Stairs}, I.~H. and {Stappers}, B.~W. and {Stinebring}, D.~R. and {Stovall}, K. and {Swiggum}, J.~K. and {Taylor}, S.~R. and {Theureau}, G. and {Tiburzi}, C. and {Toomey}, L. and {Vallisneri}, M. and {van Straten}, W. and {Vecchio}, A. and {Wang}, Y. and {Wen}, L. and {You}, X.~P. and {Zhu}, W.~W. and {Zhu}, X.-J.},
        title = "{The International Pulsar Timing Array: First data release}",
      journal = {\mnras},
         year = 2016,
        month = may,
       volume = {458},
       number = {2},
        pages = {1267-1288},
          doi = {10.1093/mnras/stw347},
archivePrefix = {arXiv},
       eprint = {1602.03640},
 primaryClass = {astro-ph.IM},
       adsurl = {https://ui.adsabs.harvard.edu/abs/2016MNRAS.458.1267V}
}

@ARTICLE{Antoniadis22,
       author = {{Antoniadis}, J. and {Arzoumanian}, Z. and {Babak}, S. and {Bailes}, M. and {Bak Nielsen}, A.-S. and {Baker}, P.~T. and {Bassa}, C.~G. and {B{\'e}csy}, B. and {Berthereau}, A. and {Bonetti}, M. and {Brazier}, A. and {Brook}, P.~R. and {Burgay}, M. and {Burke-Spolaor}, S. and {Caballero}, R.~N. and {Casey-Clyde}, J.~A. and {Chalumeau}, A. and {Champion}, D.~J. and {Charisi}, M. and {Chatterjee}, S. and {Chen}, S. and {Cognard}, I. and {Cordes}, J.~M. and {Cornish}, N.~J. and {Crawford}, F. and {Cromartie}, H.~T. and {Crowter}, K. and {Dai}, S. and {DeCesar}, M.~E. and {Demorest}, P.~B. and {Desvignes}, G. and {Dolch}, T. and {Drachler}, B. and {Falxa}, M. and {Ferrara}, E.~C. and {Fiore}, W. and {Fonseca}, E. and {Gair}, J.~R. and {Garver-Daniels}, N. and {Goncharov}, B. and {Good}, D.~C. and {Graikou}, E. and {Guillemot}, L. and {Guo}, Y.~J. and {Hazboun}, J.~S. and {Hobbs}, G. and {Hu}, H. and {Islo}, K. and {Janssen}, G.~H. and {Jennings}, R.~J. and {Johnson}, A.~D. and {Jones}, M.~L. and {Kaiser}, A.~R. and {Kaplan}, D.~L. and {Karuppusamy}, R. and {Keith}, M.~J. and {Kelley}, L.~Z. and {Kerr}, M. and {Key}, J.~S. and {Kramer}, M. and {Lam}, M.~T. and {Lamb}, W.~G. and {Lazio}, T.~J.~W. and {Lee}, K.~J. and {Lentati}, L. and {Liu}, K. and {Luo}, J. and {Lynch}, R.~S. and {Lyne}, A.~G. and {Madison}, D.~R. and {Main}, R.~A. and {Manchester}, R.~N. and {McEwen}, A. and {McKee}, J.~W. and {McLaughlin}, M.~A. and {Mickaliger}, M.~B. and {Mingarelli}, C.~M.~F. and {Ng}, C. and {Nice}, D.~J. and {Os{\l}owski}, S. and {Parthasarathy}, A. and {Pennucci}, T.~T. and {Perera}, B.~B.~P. and {Perrodin}, D. and {Petiteau}, A. and {Pol}, N.~S. and {Porayko}, N.~K. and {Possenti}, A. and {Ransom}, S.~M. and {Ray}, P.~S. and {Reardon}, D.~J. and {Russell}, C.~J. and {Samajdar}, A. and {Sampson}, L.~M. and {Sanidas}, S. and {Sarkissian}, J.~M. and {Schmitz}, K. and {Schult}, L. and {Sesana}, A. and {Shaifullah}, G. and {Shannon}, R.~M. and {Shapiro-Albert}, B.~J. and {Siemens}, X. and {Simon}, J. and {Smith}, T.~L. and {Speri}, L. and {Spiewak}, R. and {Stairs}, I.~H. and {Stappers}, B.~W. and {Stinebring}, D.~R. and {Swiggum}, J.~K. and {Taylor}, S.~R. and {Theureau}, G. and {Tiburzi}, C. and {Vallisneri}, M. and {van der Wateren}, E. and {Vecchio}, A. and {Verbiest}, J.~P.~W. and {Vigeland}, S.~J. and {Wahl}, H. and {Wang}, J.~B. and {Wang}, J. and {Wang}, L. and {Witt}, C.~A. and {Zhang}, S. and {Zhu}, X.~J.},
        title = "{The International Pulsar Timing Array second data release: Search for an isotropic gravitational wave background}",
      journal = {\mnras},
         year = 2022,
        month = mar,
       volume = {510},
       number = {4},
        pages = {4873-4887},
          doi = {10.1093/mnras/stab3418},
archivePrefix = {arXiv},
       eprint = {2201.03980},
 primaryClass = {astro-ph.HE},
       adsurl = {https://ui.adsabs.harvard.edu/abs/2022MNRAS.510.4873A}
}

@ARTICLE{Agazie23,
       author = {{Agazie}, Gabriella and {Anumarlapudi}, Akash and {Archibald}, Anne M. and {Arzoumanian}, Zaven and {Baker}, Paul T. and {B{\'e}csy}, Bence and {Blecha}, Laura and {Brazier}, Adam and {Brook}, Paul R. and {Burke-Spolaor}, Sarah and {Burnette}, Rand and {Case}, Robin and {Charisi}, Maria and {Chatterjee}, Shami and {Chatziioannou}, Katerina and {Cheeseboro}, Belinda D. and {Chen}, Siyuan and {Cohen}, Tyler and {Cordes}, James M. and {Cornish}, Neil J. and {Crawford}, Fronefield and {Cromartie}, H. Thankful and {Crowter}, Kathryn and {Cutler}, Curt J. and {Decesar}, Megan E. and {Degan}, Dallas and {Demorest}, Paul B. and {Deng}, Heling and {Dolch}, Timothy and {Drachler}, Brendan and {Ellis}, Justin A. and {Ferrara}, Elizabeth C. and {Fiore}, William and {Fonseca}, Emmanuel and {Freedman}, Gabriel E. and {Garver-Daniels}, Nate and {Gentile}, Peter A. and {Gersbach}, Kyle A. and {Glaser}, Joseph and {Good}, Deborah C. and {G{\"u}ltekin}, Kayhan and {Hazboun}, Jeffrey S. and {Hourihane}, Sophie and {Islo}, Kristina and {Jennings}, Ross J. and {Johnson}, Aaron D. and {Jones}, Megan L. and {Kaiser}, Andrew R. and {Kaplan}, David L. and {Kelley}, Luke Zoltan and {Kerr}, Matthew and {Key}, Joey S. and {Klein}, Tonia C. and {Laal}, Nima and {Lam}, Michael T. and {Lamb}, William G. and {Lazio}, T. Joseph W. and {Lewandowska}, Natalia and {Littenberg}, Tyson B. and {Liu}, Tingting and {Lommen}, Andrea and {Lorimer}, Duncan R. and {Luo}, Jing and {Lynch}, Ryan S. and {Ma}, Chung-Pei and {Madison}, Dustin R. and {Mattson}, Margaret A. and {McEwen}, Alexander and {McKee}, James W. and {McLaughlin}, Maura A. and {McMann}, Natasha and {Meyers}, Bradley W. and {Meyers}, Patrick M. and {Mingarelli}, Chiara M.~F. and {Mitridate}, Andrea and {Natarajan}, Priyamvada and {Ng}, Cherry and {Nice}, David J. and {Ocker}, Stella Koch and {Olum}, Ken D. and {Pennucci}, Timothy T. and {Perera}, Benetge B.~P. and {Petrov}, Polina and {Pol}, Nihan S. and {Radovan}, Henri A. and {Ransom}, Scott M. and {Ray}, Paul S. and {Romano}, Joseph D. and {Sardesai}, Shashwat C. and {Schmiedekamp}, Ann and {Schmiedekamp}, Carl and {Schmitz}, Kai and {Schult}, Levi and {Shapiro-Albert}, Brent J. and {Siemens}, Xavier and {Simon}, Joseph and {Siwek}, Magdalena S. and {Stairs}, Ingrid H. and {Stinebring}, Daniel R. and {Stovall}, Kevin and {Sun}, Jerry P. and {Susobhanan}, Abhimanyu and {Swiggum}, Joseph K. and {Taylor}, Jacob and {Taylor}, Stephen R. and {Turner}, Jacob E. and {Unal}, Caner and {Vallisneri}, Michele and {van Haasteren}, Rutger and {Vigeland}, Sarah J. and {Wahl}, Haley M. and {Wang}, Qiaohong and {Witt}, Caitlin A. and {Young}, Olivia and {Nanograv Collaboration}},
        title = "{The NANOGrav 15 yr Data Set: Evidence for a Gravitational-wave Background}",
      journal = {\apjl},
         year = 2023,
        month = jul,
       volume = {951},
       number = {1},
          eid = {L8},
        pages = {L8},
          doi = {10.3847/2041-8213/acdac6},
archivePrefix = {arXiv},
       eprint = {2306.16213},
 primaryClass = {astro-ph.HE},
       adsurl = {https://ui.adsabs.harvard.edu/abs/2023ApJ...951L...8A}
}

@ARTICLE{Antoniadis23,
       author = {{EPTA Collaboration} and {InPTA Collaboration} and {Antoniadis}, J. and {Arumugam}, P. and {Arumugam}, S. and {Babak}, S. and {Bagchi}, M. and {Bak Nielsen}, A.-S. and {Bassa}, C.~G. and {Bathula}, A. and {Berthereau}, A. and {Bonetti}, M. and {Bortolas}, E. and {Brook}, P.~R. and {Burgay}, M. and {Caballero}, R.~N. and {Chalumeau}, A. and {Champion}, D.~J. and {Chanlaridis}, S. and {Chen}, S. and {Cognard}, I. and {Dandapat}, S. and {Deb}, D. and {Desai}, S. and {Desvignes}, G. and {Dhanda-Batra}, N. and {Dwivedi}, C. and {Falxa}, M. and {Ferdman}, R.~D. and {Franchini}, A. and {Gair}, J.~R. and {Goncharov}, B. and {Gopakumar}, A. and {Graikou}, E. and {Grie{\ss}meier}, J.-M. and {Guillemot}, L. and {Guo}, Y.~J. and {Gupta}, Y. and {Hisano}, S. and {Hu}, H. and {Iraci}, F. and {Izquierdo-Villalba}, D. and {Jang}, J. and {Jawor}, J. and {Janssen}, G.~H. and {Jessner}, A. and {Joshi}, B.~C. and {Kareem}, F. and {Karuppusamy}, R. and {Keane}, E.~F. and {Keith}, M.~J. and {Kharbanda}, D. and {Kikunaga}, T. and {Kolhe}, N. and {Kramer}, M. and {Krishnakumar}, M.~A. and {Lackeos}, K. and {Lee}, K.~J. and {Liu}, K. and {Liu}, Y. and {Lyne}, A.~G. and {McKee}, J.~W. and {Maan}, Y. and {Main}, R.~A. and {Mickaliger}, M.~B. and {Ni{\c{t}}u}, I.~C. and {Nobleson}, K. and {Paladi}, A.~K. and {Parthasarathy}, A. and {Perera}, B.~B.~P. and {Perrodin}, D. and {Petiteau}, A. and {Porayko}, N.~K. and {Possenti}, A. and {Prabu}, T. and {Quelquejay Leclere}, H. and {Rana}, P. and {Samajdar}, A. and {Sanidas}, S.~A. and {Sesana}, A. and {Shaifullah}, G. and {Singha}, J. and {Speri}, L. and {Spiewak}, R. and {Srivastava}, A. and {Stappers}, B.~W. and {Surnis}, M. and {Susarla}, S.~C. and {Susobhanan}, A. and {Takahashi}, K. and {Tarafdar}, P. and {Theureau}, G. and {Tiburzi}, C. and {van der Wateren}, E. and {Vecchio}, A. and {Venkatraman Krishnan}, V. and {Verbiest}, J.~P.~W. and {Wang}, J. and {Wang}, L. and {Wu}, Z.},
        title = "{The second data release from the European Pulsar Timing Array. III. Search for gravitational wave signals}",
      journal = {\aap},
         year = 2023,
        month = oct,
       volume = {678},
          eid = {A50},
        pages = {A50},
          doi = {10.1051/0004-6361/202346844},
archivePrefix = {arXiv},
       eprint = {2306.16214},
 primaryClass = {astro-ph.HE},
       adsurl = {https://ui.adsabs.harvard.edu/abs/2023A&A...678A..50E}
}

@ARTICLE{Reardon23,
       author = {{Reardon}, Daniel J. and {Zic}, Andrew and {Shannon}, Ryan M. and {Hobbs}, George B. and {Bailes}, Matthew and {Di Marco}, Valentina and {Kapur}, Agastya and {Rogers}, Axl F. and {Thrane}, Eric and {Askew}, Jacob and {Bhat}, N.~D. Ramesh and {Cameron}, Andrew and {Cury{\l}o}, Ma{\l}gorzata and {Coles}, William A. and {Dai}, Shi and {Goncharov}, Boris and {Kerr}, Matthew and {Kulkarni}, Atharva and {Levin}, Yuri and {Lower}, Marcus E. and {Manchester}, Richard N. and {Mandow}, Rami and {Miles}, Matthew T. and {Nathan}, Rowina S. and {Os{\l}owski}, Stefan and {Russell}, Christopher J. and {Spiewak}, Ren{\'e}e and {Zhang}, Songbo and {Zhu}, Xing-Jiang},
        title = "{Search for an Isotropic Gravitational-wave Background with the Parkes Pulsar Timing Array}",
      journal = {\apjl},
         year = 2023,
        month = jul,
       volume = {951},
       number = {1},
          eid = {L6},
        pages = {L6},
          doi = {10.3847/2041-8213/acdd02},
archivePrefix = {arXiv},
       eprint = {2306.16215},
 primaryClass = {astro-ph.HE},
       adsurl = {https://ui.adsabs.harvard.edu/abs/2023ApJ...951L...6R}
}

@ARTICLE{Xu23,
       author = {{Xu}, Heng and {Chen}, Siyuan and {Guo}, Yanjun and {Jiang}, Jinchen and {Wang}, Bojun and {Xu}, Jiangwei and {Xue}, Zihan and {Caballero}, R. Nicolas and {Yuan}, Jianping and {Xu}, Yonghua and {Wang}, Jingbo and {Hao}, Longfei and {Luo}, Jingtao and {Lee}, Kejia and {Han}, Jinlin and {Jiang}, Peng and {Shen}, Zhiqiang and {Wang}, Min and {Wang}, Na and {Xu}, Renxin and {Wu}, Xiangping and {Manchester}, Richard and {Qian}, Lei and {Guan}, Xin and {Huang}, Menglin and {Sun}, Chun and {Zhu}, Yan},
        title = "{Searching for the Nano-Hertz Stochastic Gravitational Wave Background with the Chinese Pulsar Timing Array Data Release I}",
      journal = {Research in Astronomy and Astrophysics},
         year = 2023,
        month = jul,
       volume = {23},
       number = {7},
          eid = {075024},
        pages = {075024},
          doi = {10.1088/1674-4527/acdfa5},
archivePrefix = {arXiv},
       eprint = {2306.16216},
 primaryClass = {astro-ph.HE},
       adsurl = {https://ui.adsabs.harvard.edu/abs/2023RAA....23g5024X}
}

@ARTICLE{Taylor89,
       author = {{Taylor}, J.~H. and {Weisberg}, J.~M.},
        title = "{Further Experimental Tests of Relativistic Gravity Using the Binary Pulsar PSR 1913+16}",
      journal = {\apj},
         year = 1989,
        month = oct,
       volume = {345},
        pages = {434},
          doi = {10.1086/167917},
       adsurl = {https://ui.adsabs.harvard.edu/abs/1989ApJ...345..434T}
}

@ARTICLE{vanStraten11,
       author = {{van Straten}, W. and {Bailes}, M.},
        title = "{DSPSR: Digital Signal Processing Software for Pulsar Astronomy}",
      journal = {\pasa},
         year = 2011,
        month = jan,
       volume = {28},
       number = {1},
        pages = {1-14},
          doi = {10.1071/AS10021},
archivePrefix = {arXiv},
       eprint = {1008.3973},
 primaryClass = {astro-ph.IM},
       adsurl = {https://ui.adsabs.harvard.edu/abs/2011PASA...28....1V}
}

@ARTICLE{vanStraten12,
       author = {{van Straten}, Willem and {Demorest}, Paul and {Oslowski}, Stefan},
        title = "{Pulsar Data Analysis with PSRCHIVE}",
      journal = {Astronomical Research and Technology},
         year = 2012,
        month = jul,
       volume = {9},
       number = {3},
        pages = {237-256},
archivePrefix = {arXiv},
       eprint = {1205.6276},
 primaryClass = {astro-ph.IM},
       adsurl = {https://ui.adsabs.harvard.edu/abs/2012AR&T....9..237V}
}

@ARTICLE{Hotan04,
       author = {{Hotan}, A.~W. and {van Straten}, W. and {Manchester}, R.~N.},
        title = "{PSRCHIVE and PSRFITS: An Open Approach to Radio Pulsar Data Storage and Analysis}",
      journal = {\pasa},
         year = 2004,
        month = jan,
       volume = {21},
       number = {3},
        pages = {302-309},
          doi = {10.1071/AS04022},
archivePrefix = {arXiv},
       eprint = {astro-ph/0404549},
 primaryClass = {astro-ph},
       adsurl = {https://ui.adsabs.harvard.edu/abs/2004PASA...21..302H}
}

@ARTICLE{Jiang20,
       author = {{Jiang}, Peng and {Tang}, Ning-Yu and {Hou}, Li-Gang and {Liu}, Meng-Ting and {Kr{\v{c}}o}, Marko and {Qian}, Lei and {Sun}, Jing-Hai and {Ching}, Tao-Chung and {Liu}, Bin and {Duan}, Yan and {Yue}, You-Ling and {Gan}, Heng-Qian and {Yao}, Rui and {Li}, Hui and {Pan}, Gao-Feng and {Yu}, Dong-Jun and {Liu}, Hong-Fei and {Li}, Di and {Peng}, Bo and {Yan}, Jun and {FAST Collaboration}},
        title = "{The fundamental performance of FAST with 19-beam receiver at L band}",
      journal = {Research in Astronomy and Astrophysics},
         year = 2020,
        month = may,
       volume = {20},
       number = {5},
          eid = {064},
        pages = {064},
          doi = {10.1088/1674-4527/20/5/64},
archivePrefix = {arXiv},
       eprint = {2002.01786},
 primaryClass = {astro-ph.IM},
       adsurl = {https://ui.adsabs.harvard.edu/abs/2020RAA....20...64J}
}

@ARTICLE{McLaughlin06,
       author = {{McLaughlin}, M.~A. and {Lyne}, A.~G. and {Lorimer}, D.~R. and {Kramer}, M. and {Faulkner}, A.~J. and {Manchester}, R.~N. and {Cordes}, J.~M. and {Camilo}, F. and {Possenti}, A. and {Stairs}, I.~H. and {Hobbs}, G. and {D'Amico}, N. and {Burgay}, M. and {O'Brien}, J.~T.},
        title = "{Transient radio bursts from rotating neutron stars}",
      journal = {\nat},
         year = 2006,
        month = feb,
       volume = {439},
       number = {7078},
        pages = {817-820},
          doi = {10.1038/nature04440},
archivePrefix = {arXiv},
       eprint = {astro-ph/0511587},
 primaryClass = {astro-ph},
       adsurl = {https://ui.adsabs.harvard.edu/abs/2006Natur.439..817M}
}

@ARTICLE{Keane08,
       author = {{Keane}, E.~F. and {Kramer}, M.},
        title = "{On the birthrates of Galactic neutron stars}",
      journal = {\mnras},
         year = 2008,
        month = dec,
       volume = {391},
       number = {4},
        pages = {2009-2016},
          doi = {10.1111/j.1365-2966.2008.14045.x},
archivePrefix = {arXiv},
       eprint = {0810.1512},
 primaryClass = {astro-ph},
       adsurl = {https://ui.adsabs.harvard.edu/abs/2008MNRAS.391.2009K}
}

@ARTICLE{McLaughlin09,
       author = {{McLaughlin}, M.~A. and {Lyne}, A.~G. and {Keane}, E.~F. and {Kramer}, M. and {Miller}, J.~J. and {Lorimer}, D.~R. and {Manchester}, R.~N. and {Camilo}, F. and {Stairs}, I.~H.},
        title = "{Timing observations of rotating radio transients}",
      journal = {\mnras},
         year = 2009,
        month = dec,
       volume = {400},
       number = {3},
        pages = {1431-1438},
          doi = {10.1111/j.1365-2966.2009.15584.x},
archivePrefix = {arXiv},
       eprint = {0908.3813},
 primaryClass = {astro-ph.SR},
       adsurl = {https://ui.adsabs.harvard.edu/abs/2009MNRAS.400.1431M}
}

@ARTICLE{Bhattacharyya18,
       author = {{Bhattacharyya}, B. and {Lyne}, A.~G. and {Stappers}, B.~W. and {Weltevrede}, P. and {Keane}, E.~F. and {McLaughlin}, M.~A. and {Kramer}, M. and {Jordan}, C. and {Bassa}, C.},
        title = "{A long-term study of three rotating radio transients}",
      journal = {\mnras},
         year = 2018,
        month = jul,
       volume = {477},
       number = {3},
        pages = {4090-4103},
          doi = {10.1093/mnras/sty923},
archivePrefix = {arXiv},
       eprint = {1803.10277},
 primaryClass = {astro-ph.HE},
       adsurl = {https://ui.adsabs.harvard.edu/abs/2018MNRAS.477.4090B}
}

@ARTICLE{Wang07,
       author = {{Wang}, N. and {Manchester}, R.~N. and {Johnston}, S.},
        title = "{Pulsar nulling and mode changing}",
      journal = {\mnras},
         year = 2007,
        month = may,
       volume = {377},
       number = {3},
        pages = {1383-1392},
          doi = {10.1111/j.1365-2966.2007.11703.x},
archivePrefix = {arXiv},
       eprint = {astro-ph/0703241},
 primaryClass = {astro-ph},
       adsurl = {https://ui.adsabs.harvard.edu/abs/2007MNRAS.377.1383W}
}

@ARTICLE{Zhang24,
       author = {{Zhang}, S.~B. and {Geng}, J.~J. and {Wang}, J.~S. and {Yang}, X. and {Kaczmarek}, J. and {Tang}, Z.~F. and {Johnston}, S. and {Hobbs}, G. and {Manchester}, R. and {Wu}, X.~F. and {Jiang}, P. and {Huang}, Y.~F. and {Zou}, Y.~C. and {Dai}, Z.~G. and {Zhang}, B. and {Li}, D. and {Yang}, Y.~P. and {Dai}, S. and {Chang}, C.~M. and {Pan}, Z.~C. and {Lu}, J.~G. and {Wei}, J.~J. and {Li}, Y. and {Wu}, Q.~W. and {Qian}, L. and {Wang}, P. and {Wang}, S.~Q. and {Feng}, Y. and {Staveley-Smith}, L.},
        title = "{RRAT J1913+1330: An Extremely Variable and Puzzling Pulsar}",
      journal = {\apj},
         year = 2024,
        month = sep,
       volume = {972},
       number = {1},
          eid = {59},
        pages = {59},
          doi = {10.3847/1538-4357/ad6602},
archivePrefix = {arXiv},
       eprint = {2306.02855},
 primaryClass = {astro-ph.HE},
       adsurl = {https://ui.adsabs.harvard.edu/abs/2024ApJ...972...59Z}
}

@ARTICLE{Nimmo22,
       author = {{Nimmo}, K. and {Hessels}, J.~W.~T. and {Kirsten}, F. and {Keimpema}, A. and {Cordes}, J.~M. and {Snelders}, M.~P. and {Hewitt}, D.~M. and {Karuppusamy}, R. and {Archibald}, A.~M. and {Bezrukovs}, V. and {Bhardwaj}, M. and {Blaauw}, R. and {Buttaccio}, S.~T. and {Cassanelli}, T. and {Conway}, J.~E. and {Corongiu}, A. and {Feiler}, R. and {Fonseca}, E. and {Forss{\'e}n}, O. and {Gawro{\'n}ski}, M. and {Giroletti}, M. and {Kharinov}, M.~A. and {Leung}, C. and {Lindqvist}, M. and {Maccaferri}, G. and {Marcote}, B. and {Masui}, K.~W. and {Mckinven}, R. and {Melnikov}, A. and {Michilli}, D. and {Mikhailov}, A.~G. and {Ng}, C. and {Orbidans}, A. and {Ould-Boukattine}, O.~S. and {Paragi}, Z. and {Pearlman}, A.~B. and {Petroff}, E. and {Rahman}, M. and {Scholz}, P. and {Shin}, K. and {Smith}, K.~M. and {Stairs}, I.~H. and {Surcis}, G. and {Tendulkar}, S.~P. and {Vlemmings}, W. and {Wang}, N. and {Yang}, J. and {Yuan}, J.~P.},
        title = "{Burst timescales and luminosities as links between young pulsars and fast radio bursts}",
      journal = {Nature Astronomy},
         year = 2022,
        month = feb,
       volume = {6},
        pages = {393-401},
          doi = {10.1038/s41550-021-01569-9},
archivePrefix = {arXiv},
       eprint = {2105.11446},
 primaryClass = {astro-ph.HE},
       adsurl = {https://ui.adsabs.harvard.edu/abs/2022NatAs...6..393N}
}

@PHDTHESIS{Ransom01,
       author = {{Ransom}, Scott Mitchell},
        title = "{New search techniques for binary pulsars}",
       school = {Harvard University, Massachusetts},
         year = 2001,
        month = jan,
       adsurl = {https://ui.adsabs.harvard.edu/abs/2001PhDT.......123R}
}

@article{Branch99,
author = {Branch, Mary Ann and Coleman, Thomas F. and Li, Yuying},
title = {A Subspace, Interior, and Conjugate Gradient Method for Large-Scale Bound-Constrained Minimization Problems},
journal = {SIAM Journal on Scientific Computing},
volume = {21},
number = {1},
pages = {1-23},
year = {1999},
doi = {10.1137/S1064827595289108},
URL = { https://doi.org/10.1137/S1064827595289108},
eprint = { https://doi.org/10.1137/S1064827595289108}
}

@ARTICLE{Ashton20,
       author = {{Ashton}, Gregory and {Lasky}, Paul D. and {Nathan}, Rowina and {Palfreyman}, Jim},
        title = "{Flickering of the Vela pulsar during its 2016 glitch}",
      journal = {arXiv e-prints},
         year = 2020,
        month = nov,
          eid = {arXiv:2011.07927},
        pages = {arXiv:2011.07927},
          doi = {10.48550/arXiv.2011.07927},
archivePrefix = {arXiv},
       eprint = {2011.07927},
 primaryClass = {astro-ph.HE},
       adsurl = {https://ui.adsabs.harvard.edu/abs/2020arXiv201107927A}
}

@ARTICLE{Refregier03,
       author = {{Refregier}, Alexandre},
        title = "{Shapelets - I. A method for image analysis}",
      journal = {\mnras},
         year = 2003,
        month = jan,
       volume = {338},
       number = {1},
        pages = {35-47},
          doi = {10.1046/j.1365-8711.2003.05901.x},
archivePrefix = {arXiv},
       eprint = {astro-ph/0105178},
 primaryClass = {astro-ph},
       adsurl = {https://ui.adsabs.harvard.edu/abs/2003MNRAS.338...35R}
}

@ARTICLE{emcee13,
       author = {{Foreman-Mackey}, Daniel and {Hogg}, David W. and {Lang}, Dustin and {Goodman}, Jonathan},
        title = "{emcee: The MCMC Hammer}",
      journal = {\pasp},
         year = 2013,
        month = mar,
       volume = {125},
       number = {925},
        pages = {306},
          doi = {10.1086/670067},
archivePrefix = {arXiv},
       eprint = {1202.3665},
 primaryClass = {astro-ph.IM},
       adsurl = {https://ui.adsabs.harvard.edu/abs/2013PASP..125..306F}
}

@ARTICLE{Akaike74,
  author={Akaike, H.},
  journal={IEEE Transactions on Automatic Control}, 
  title={A new look at the statistical model identification}, 
  year={1974},
  volume={19},
  number={6},
  pages={716-723},
  doi={10.1109/TAC.1974.1100705}
}

@ARTICLE{Hobbs06,
       author = {{Hobbs}, G.~B. and {Edwards}, R.~T. and {Manchester}, R.~N.},
        title = "{TEMPO2, a new pulsar-timing package - I. An overview}",
      journal = {\mnras},
         year = 2006,
        month = jun,
       volume = {369},
       number = {2},
        pages = {655-672},
          doi = {10.1111/j.1365-2966.2006.10302.x},
archivePrefix = {arXiv},
       eprint = {astro-ph/0603381},
 primaryClass = {astro-ph},
       adsurl = {https://ui.adsabs.harvard.edu/abs/2006MNRAS.369..655H}
}

@ARTICLE{Edwards06,
       author = {{Edwards}, R.~T. and {Hobbs}, G.~B. and {Manchester}, R.~N.},
        title = "{TEMPO2, a new pulsar timing package - II. The timing model and precision estimates}",
      journal = {\mnras},
         year = 2006,
        month = nov,
       volume = {372},
       number = {4},
        pages = {1549-1574},
          doi = {10.1111/j.1365-2966.2006.10870.x},
archivePrefix = {arXiv},
       eprint = {astro-ph/0607664},
 primaryClass = {astro-ph},
       adsurl = {https://ui.adsabs.harvard.edu/abs/2006MNRAS.372.1549E}
}

@ARTICLE{Li26,
       author = {{Li}, Y. and {Zhang}, S.~B. and {Yang}, Y.~P. and {Tsai}, C.~W. and {Yang}, X. and {Law}, C.~J. and {Anna-Thomas}, R. and {Chen}, X.~L. and {Lee}, K.~J. and {Tang}, Z.~F. and {Xiao}, D. and {Xu}, H. and {Yang}, X.~L. and {Chen}, G. and {Feng}, Y. and {Li}, D.~Z. and {Mckinven}, R. and {Niu}, J.~R. and {Shin}, K. and {Wang}, B.~J. and {Zhang}, C.~F. and {Zhang}, Y.~K. and {Zhou}, D.~J. and {Zhu}, Y.~H. and {Dai}, Z.~G. and {Chang}, C.~M. and {Geng}, J.~J. and {Han}, J.~L. and {Hu}, L. and {Li}, D. and {Luo}, R. and {Niu}, C.~H. and {Shi}, D.~D. and {Sun}, T.~R. and {Wu}, X.~F. and {Zhu}, W.~W. and {Jiang}, P. and {Zhang}, B.},
        title = "{A sudden change and recovery in the magnetic environment around a repeating fast radio burst}",
      journal = {Science},
         year = 2026,
        month = jan,
       volume = {391},
       number = {6782},
        pages = {280-284},
          doi = {10.1126/science.adq3225},
archivePrefix = {arXiv},
       eprint = {2503.04727},
 primaryClass = {astro-ph.HE},
       adsurl = {https://ui.adsabs.harvard.edu/abs/2026Sci...391..280L}
}

@ARTICLE{Hickish16,
       author = {{Hickish}, Jack and {Abdurashidova}, Zuhra and {Ali}, Zaki and {Buch}, Kaushal D. and {Chaudhari}, Sandeep C. and {Chen}, Hong and {Dexter}, Matthew and {Domagalski}, Rachel Simone and {Ford}, John and {Foster}, Griffin and {George}, David and {Greenberg}, Joe and {Greenhill}, Lincoln and {Isaacson}, Adam and {Jiang}, Homin and {Jones}, Glenn and {Kapp}, Francois and {Kriel}, Henno and {Lacasse}, Rich and {Lutomirski}, Andrew and {MacMahon}, David and {Manley}, Jason and {Martens}, Andrew and {McCullough}, Randy and {Muley}, Mekhala V. and {New}, Wesley and {Parsons}, Aaron and {Price}, Daniel C. and {Primiani}, Rurik A. and {Ray}, Jason and {Siemion}, Andrew and {van Tonder}, Verees{\'e} and {Vertatschitsch}, Laura and {Wagner}, Mark and {Weintroub}, Jonathan and {Werthimer}, Dan},
        title = "{A Decade of Developing Radio-Astronomy Instrumentation using CASPER Open-Source Technology}",
      journal = {Journal of Astronomical Instrumentation},
         year = 2016,
        month = dec,
       volume = {5},
       number = {4},
          eid = {1641001-12},
        pages = {1641001-12},
          doi = {10.1142/S2251171716410014},
archivePrefix = {arXiv},
       eprint = {1611.01826},
 primaryClass = {astro-ph.IM},
       adsurl = {https://ui.adsabs.harvard.edu/abs/2016JAI.....541001H}
}

@ARTICLE{Li18,
       author = {{Li}, Di and {Wang}, Pei and {Qian}, Lei and {Krco}, Marko and {Jiang}, Peng and {Yue}, Youling and {Jin}, Chenjin and {Zhu}, Yan and {Pan}, Zhichen and {Nan}, Rendong and {Dunning}, Alex},
        title = "{FAST in Space: Considerations for a Multibeam, Multipurpose Survey Using China's 500-m Aperture Spherical Radio Telescope (FAST)}",
      journal = {IEEE Microwave Magazine},
         year = 2018,
        month = apr,
       volume = {19},
       number = {3},
        pages = {112-119},
          doi = {10.1109/MMM.2018.2802178},
archivePrefix = {arXiv},
       eprint = {1802.03709},
 primaryClass = {astro-ph.IM},
       adsurl = {https://ui.adsabs.harvard.edu/abs/2018IMMag..19..112L}
}

@ARTICLE{Kimpson20,
       author = {{Kimpson}, Tom and {Wu}, Kinwah and {Zane}, Silvia},
        title = "{Orbital spin dynamics of a millisecond pulsar around a massive BH with a general mass quadrupole}",
      journal = {\mnras},
         year = 2020,
        month = oct,
       volume = {497},
       number = {4},
        pages = {5421-5431},
          doi = {10.1093/mnras/staa2103},
archivePrefix = {arXiv},
       eprint = {2007.05219},
 primaryClass = {astro-ph.HE},
       adsurl = {https://ui.adsabs.harvard.edu/abs/2020MNRAS.497.5421K}
}

@ARTICLE{Xiao24,
       author = {{Xiao}, Xinxu and {Shen}, Rong-Feng},
        title = "{Apparently Ultralong Period Radio Signals from Self-lensed Pulsar─Black Hole Binaries}",
      journal = {\apj},
         year = 2024,
        month = sep,
       volume = {972},
       number = {1},
          eid = {60},
        pages = {60},
          doi = {10.3847/1538-4357/ad65d8},
archivePrefix = {arXiv},
       eprint = {2401.12494},
 primaryClass = {astro-ph.HE},
       adsurl = {https://ui.adsabs.harvard.edu/abs/2024ApJ...972...60X}
}

@ARTICLE{Johnston06,
       author = {{Johnston}, Simon and {Kramer}, M. and {Lorimer}, D.~R. and {Lyne}, A.~G. and {McLaughlin}, M. and {Klein}, B. and {Manchester}, R.~N.},
        title = "{Discovery of two pulsars towards the Galactic Centre}",
      journal = {\mnras},
         year = 2006,
        month = nov,
       volume = {373},
       number = {1},
        pages = {L6-L10},
          doi = {10.1111/j.1745-3933.2006.00232.x},
archivePrefix = {arXiv},
       eprint = {astro-ph/0606465},
 primaryClass = {astro-ph},
       adsurl = {https://ui.adsabs.harvard.edu/abs/2006MNRAS.373L...6J}
}

@ARTICLE{astropy22,
       author = {{Astropy Collaboration} and {Price-Whelan}, Adrian M. and {Lim}, Pey Lian and {Earl}, Nicholas and {Starkman}, Nathaniel and {Bradley}, Larry and {Shupe}, David L. and {Patil}, Aarya A. and {Corrales}, Lia and {Brasseur}, C.~E. and {N{\"o}the}, Maximilian and {Donath}, Axel and {Tollerud}, Erik and {Morris}, Brett M. and {Ginsburg}, Adam and {Vaher}, Eero and {Weaver}, Benjamin A. and {Tocknell}, James and {Jamieson}, William and {van Kerkwijk}, Marten H. and {Robitaille}, Thomas P. and {Merry}, Bruce and {Bachetti}, Matteo and {G{\"u}nther}, H. Moritz and {Aldcroft}, Thomas L. and {Alvarado-Montes}, Jaime A. and {Archibald}, Anne M. and {B{\'o}di}, Attila and {Bapat}, Shreyas and {Barentsen}, Geert and {Baz{\'a}n}, Juanjo and {Biswas}, Manish and {Boquien}, M{\'e}d{\'e}ric and {Burke}, D.~J. and {Cara}, Daria and {Cara}, Mihai and {Conroy}, Kyle E. and {Conseil}, Simon and {Craig}, Matthew W. and {Cross}, Robert M. and {Cruz}, Kelle L. and {D'Eugenio}, Francesco and {Dencheva}, Nadia and {Devillepoix}, Hadrien A.~R. and {Dietrich}, J{\"o}rg P. and {Eigenbrot}, Arthur Davis and {Erben}, Thomas and {Ferreira}, Leonardo and {Foreman-Mackey}, Daniel and {Fox}, Ryan and {Freij}, Nabil and {Garg}, Suyog and {Geda}, Robel and {Glattly}, Lauren and {Gondhalekar}, Yash and {Gordon}, Karl D. and {Grant}, David and {Greenfield}, Perry and {Groener}, Austen M. and {Guest}, Steve and {Gurovich}, Sebastian and {Handberg}, Rasmus and {Hart}, Akeem and {Hatfield-Dodds}, Zac and {Homeier}, Derek and {Hosseinzadeh}, Griffin and {Jenness}, Tim and {Jones}, Craig K. and {Joseph}, Prajwel and {Kalmbach}, J. Bryce and {Karamehmetoglu}, Emir and {Ka{\l}uszy{\'n}ski}, Miko{\l}aj and {Kelley}, Michael S.~P. and {Kern}, Nicholas and {Kerzendorf}, Wolfgang E. and {Koch}, Eric W. and {Kulumani}, Shankar and {Lee}, Antony and {Ly}, Chun and {Ma}, Zhiyuan and {MacBride}, Conor and {Maljaars}, Jakob M. and {Muna}, Demitri and {Murphy}, N.~A. and {Norman}, Henrik and {O'Steen}, Richard and {Oman}, Kyle A. and {Pacifici}, Camilla and {Pascual}, Sergio and {Pascual-Granado}, J. and {Patil}, Rohit R. and {Perren}, Gabriel I. and {Pickering}, Timothy E. and {Rastogi}, Tanuj and {Roulston}, Benjamin R. and {Ryan}, Daniel F. and {Rykoff}, Eli S. and {Sabater}, Jose and {Sakurikar}, Parikshit and {Salgado}, Jes{\'u}s and {Sanghi}, Aniket and {Saunders}, Nicholas and {Savchenko}, Volodymyr and {Schwardt}, Ludwig and {Seifert-Eckert}, Michael and {Shih}, Albert Y. and {Jain}, Anany Shrey and {Shukla}, Gyanendra and {Sick}, Jonathan and {Simpson}, Chris and {Singanamalla}, Sudheesh and {Singer}, Leo P. and {Singhal}, Jaladh and {Sinha}, Manodeep and {Sip{\H{o}}cz}, Brigitta M. and {Spitler}, Lee R. and {Stansby}, David and {Streicher}, Ole and {{\v{S}}umak}, Jani and {Swinbank}, John D. and {Taranu}, Dan S. and {Tewary}, Nikita and {Tremblay}, Grant R. and {de Val-Borro}, Miguel and {Van Kooten}, Samuel J. and {Vasovi{\'c}}, Zlatan and {Verma}, Shresth and {de Miranda Cardoso}, Jos{\'e} Vin{\'\i}cius and {Williams}, Peter K.~G. and {Wilson}, Tom J. and {Winkel}, Benjamin and {Wood-Vasey}, W.~M. and {Xue}, Rui and {Yoachim}, Peter and {Zhang}, Chen and {Zonca}, Andrea and {Astropy Project Contributors}},
        title = "{The Astropy Project: Sustaining and Growing a Community-oriented Open-source Project and the Latest Major Release (v5.0) of the Core Package}",
      journal = {\apj},
         year = 2022,
        month = aug,
       volume = {935},
       number = {2},
          eid = {167},
        pages = {167},
          doi = {10.3847/1538-4357/ac7c74},
archivePrefix = {arXiv},
       eprint = {2206.14220},
 primaryClass = {astro-ph.IM},
       adsurl = {https://ui.adsabs.harvard.edu/abs/2022ApJ...935..167A}
}

@ARTICLE{scipy20,
       author = {{Virtanen}, Pauli and {Gommers}, Ralf and {Oliphant}, Travis E. and {Haberland}, Matt and {Reddy}, Tyler and {Cournapeau}, David and {Burovski}, Evgeni and {Peterson}, Pearu and {Weckesser}, Warren and {Bright}, Jonathan and {van der Walt}, St{\'e}fan J. and {Brett}, Matthew and {Wilson}, Joshua and {Millman}, K. Jarrod and {Mayorov}, Nikolay and {Nelson}, Andrew R.~J. and {Jones}, Eric and {Kern}, Robert and {Larson}, Eric and {Carey}, C.~J. and {Polat}, {\.I}lhan and {Feng}, Yu and {Moore}, Eric W. and {VanderPlas}, Jake and {Laxalde}, Denis and {Perktold}, Josef and {Cimrman}, Robert and {Henriksen}, Ian and {Quintero}, E.~A. and {Harris}, Charles R. and {Archibald}, Anne M. and {Ribeiro}, Ant{\^o}nio H. and {Pedregosa}, Fabian and {van Mulbregt}, Paul and {SciPy 1.  0 Contributors}},
        title = "{SciPy 1.0: fundamental algorithms for scientific computing in Python}",
      journal = {Nature Medicine},
         year = 2020,
        month = feb,
       volume = {17},
        pages = {261-272},
          doi = {10.1038/s41592-019-0686-2},
archivePrefix = {arXiv},
       eprint = {1907.10121},
 primaryClass = {cs.MS},
       adsurl = {https://ui.adsabs.harvard.edu/abs/2020NaMet..17..261V}
}

@ARTICLE{numpy20,
       author = {{Harris}, Charles R. and {Millman}, K. Jarrod and {van der Walt}, St{\'e}fan J. and {Gommers}, Ralf and {Virtanen}, Pauli and {Cournapeau}, David and {Wieser}, Eric and {Taylor}, Julian and {Berg}, Sebastian and {Smith}, Nathaniel J. and {Kern}, Robert and {Picus}, Matti and {Hoyer}, Stephan and {van Kerkwijk}, Marten H. and {Brett}, Matthew and {Haldane}, Allan and {del R{\'\i}o}, Jaime Fern{\'a}ndez and {Wiebe}, Mark and {Peterson}, Pearu and {G{\'e}rard-Marchant}, Pierre and {Sheppard}, Kevin and {Reddy}, Tyler and {Weckesser}, Warren and {Abbasi}, Hameer and {Gohlke}, Christoph and {Oliphant}, Travis E.},
        title = "{Array programming with NumPy}",
      journal = {\nat},
         year = 2020,
        month = sep,
       volume = {585},
       number = {7825},
        pages = {357-362},
          doi = {10.1038/s41586-020-2649-2},
archivePrefix = {arXiv},
       eprint = {2006.10256},
 primaryClass = {cs.MS},
       adsurl = {https://ui.adsabs.harvard.edu/abs/2020Natur.585..357H}
}

@ARTICLE{Zhang24_highB,
       author = {{Zhang}, S.~B. and {Yang}, X. and {Geng}, J.~J. and {Yang}, Y.~P. and {Wu}, X.~F.},
        title = "{Detection of Hidden Emissions in Two Rotating Radio Transients with High Surface Magnetic Fields}",
      journal = {\apjl},
         year = 2024,
        month = dec,
       volume = {976},
       number = {2},
          eid = {L26},
        pages = {L26},
          doi = {10.3847/2041-8213/ad92fb},
archivePrefix = {arXiv},
       eprint = {2407.09876},
 primaryClass = {astro-ph.HE},
       adsurl = {https://ui.adsabs.harvard.edu/abs/2024ApJ...976L..26Z}
}
\bibliographystyle{aasjournal}

\end{document}